\documentclass[journal,twoside]{IEEEtran}

\usepackage{ifpdf}
\ifpdf
\else
  \ifdefined\XeTeXversion
  \else
    \PassOptionsToPackage{dvipdfmx}{graphicx}
    \PassOptionsToPackage{dvipdfmx}{color}
    \PassOptionsToPackage{dvipdfmx}{hyperref}
  \fi
\fi

\usepackage[T1]{fontenc}
\usepackage{algorithmic}
\usepackage{algorithm}
\usepackage{array}
\usepackage{textcomp}
\usepackage{stfloats}
\usepackage{url}
\usepackage{verbatim}
\usepackage{graphicx}
\usepackage{amsfonts}
\usepackage{amsmath}
\usepackage{amssymb}
\newtheorem{dfn}{Definition}
\newtheorem{theo}{Theorem}
\newtheorem{rem}{Remark}
\usepackage{color}
\usepackage{booktabs}
\usepackage{multirow}
\usepackage[nocompress]{cite}
\makeatletter
\let\MYcaption\@makecaption
\makeatother
\usepackage[justification=centering,labelfont=sf,textfont=sf]{subcaption}
\makeatletter
\let\@makecaption\MYcaption
\makeatother
\usepackage{setspace}
\usepackage{cellspace}
\input{10_style/newcommands.sty}
\usepackage{hyperref}
\hypersetup{
    colorlinks = true,
    urlcolor = black,
    linkcolor = black,
    citecolor = black,
    anchorcolor = black,
    pdfborder = {0 0 0}
}
\usepackage{enumitem}

\makeatletter
\@twosidetrue
\makeatother
\renewcommand{\IEEEtitletopspaceextra}{-6.127bp}
\IEEEaftertitletext{\vspace{-9.365bp}}
\hypersetup{linkcolor=blue,citecolor=blue,urlcolor=black,anchorcolor=black}
\makeatletter
\def\@IEEEinterspaceratioM{0.35}
\def\@IEEEinterspaceMINratioM{0.10}
\def\@IEEEinterspaceMINratioB{0.23}
\makeatother
\newenvironment{publicationdisplaypadding}[1]{%
  \begingroup
  \addtolength{\abovedisplayskip}{#1}%
  \addtolength{\belowdisplayskip}{#1}%
  \addtolength{\abovedisplayshortskip}{#1}%
  \addtolength{\belowdisplayshortskip}{#1}%
}{\par\endgroup}

\newcommand{\PublicationFlushColumn}{\par\vskip-\lastskip\penalty-10000\relax}
\newcommand{\PublicationFlushPage}{\PublicationFlushColumn\clearpage}
\newcommand{\PublicationFlushDisplayColumn}[1]{\par\vskip-\lastskip\kern-#1\penalty-10000\relax}

\input{10_style/publication-panels.tex}

\begin{document}
\title{Generalized Graph Signal Sampling by Difference-of-Convex Optimization}

\author{Keitaro~Yamashita,~\IEEEmembership{Member,~IEEE,}
        Kazuki~Naganuma,~\IEEEmembership{Member,~IEEE,}\\
        and~Shunsuke~Ono,~\IEEEmembership{Senior Member,~IEEE}%
\thanks{This is the author-prepared version of an article published in \emph{IEEE Transactions on Signal Processing}, vol.~74, pp.~4653--4667, 2026, doi: \href{https://doi.org/10.1109/TSP.2026.3741113}{10.1109/TSP.2026.3741113}.}
\thanks{Received 22 June 2026; revised 27 September 2026; accepted 2 October 2026. Date of publication 7 October 2026; date of current version 23 October 2026. This work was supported in part by JST FOREST under Grant JPMJFR232M; in part by JST CREST under Grant JPMJCR25Q5; in part by JST ACT-X under Grant JPMJAX23CJ; and in part by JSPS KAKENHI
under Grant 24K22291, Grant 25H01296, Grant 25K03136, Grant 26H02536, Grant 25KJ0117, and Grant 26K21246. The associate editor coordinating the review of this article and approving it for publication was Hoi-To Wai. (Corresponding author: Keitaro Yamashita.)}
\thanks{Keitaro Yamashita and Shunsuke Ono are with the Department of Computer Science, Institute of Science Tokyo, Tokyo 152-8550, Japan (e-mail: yamashita@mdi.comp.isct.ac.jp; ono@comp.isct.ac.jp).}
\thanks{Kazuki Naganuma is with the Institute of Engineering, Tokyo University of Agriculture and Technology, Tokyo 183-8538, Japan (e-mail: k-naganuma@go.tuat.ac.jp).}
\thanks{Digital Object Identifier 10.1109/TSP.2026.3741113}}

\markboth{Author-prepared version, IEEE Transactions on Signal Processing, vol.~74, 2026}{Yamashita \MakeLowercase{\textit{et al.}}: Generalized Graph Signal Sampling by Difference-of-Convex Optimization}

\maketitle
\enlargethispage{-36bp}

\widowpenalty=10000
\clubpenalty=10000
\begin{abstract}
    We propose a comprehensive framework for the generalized sampling and recovery of graph signals by leveraging difference-of-convex (DC) optimization.
    A fundamental challenge in graph signal processing is sampling, especially for graph signals that are not bandlimited.
    To accurately capture complex real-world phenomena, it is essential to handle beyond bandlimited graph signals, moving past traditional bandlimited assumptions.
    Consequently, extending the generalized sampling theory to graph signals has been studied, enabling the best possible recovery for a wide range of signals by assuming signal priors.
    However, achieving the best possible recovery requires handling inherently non-convex and computationally intractable constraints such as the full rank constraint.
    As a result, existing methods have relied on either aggressive convex relaxations that sacrifice accuracy or greedy algorithms that risk falling into poor suboptimal solutions, facing a fundamental dilemma between modeling accuracy and optimization tractability.
    To overcome this dilemma, we propose a DC optimization-based method for designing an aggregation sampling operator for beyond bandlimited graph signals that comprehensively handles arbitrary signal priors assumed in the generalized sampling theory.
    Specifically, the intractable full rank constraint is relaxed using the nuclear norm, reformulating the design problem into a DC optimization problem.
    We develop a solver based on the general double-proximal gradient DC algorithm, which theoretically guarantees convergence to a critical point.
    Experimental results on synthetic and real-world data demonstrate the superiority of our method in sampling and recovering beyond bandlimited graph signals compared to existing approaches (see \url{https://www.mdi.comp.isct.ac.jp}).
\end{abstract}


\IEEEpeerreviewmaketitle
\bstctlcite{arxiv:bstcontrol}

\newpage
\enlargethispage{-36bp}
\section{Introduction}
\label{sec:intro}

\IEEEPARstart{G}{raph} signal processing (GSP) presents an innovative approach to analyzing data depicted as signals situated at the vertices of a graph.
GSP is an active and dynamic domain of investigation within signal processing, encompassing various inquiries ranging from theoretical investigations to real-world implementations.
Recent research in GSP has focused on extending classical signal processing theories to the graph setting~\cite{narang2012perfect,agaskar2013spectral,sandryhaila2014big,shuman2015multiscale,tanaka2020gensamp}.
There are several aspects of research related to GSP.
Typical examples include graph learning~\cite{dong2016learning,egilmez2018graph,dong2019learning}
and graph signal recovery~\cite{ono2015total,onuki2016graph,nagahama2022graph,yamagata2025robust,dabush2026efficient}.
There are also various applications of GSP, including image and point cloud processing~\cite{cheung2018graph,wang2019dynamic,takemoto2022graph}, graph neural networks~\cite{bronstein2017geometric}, machine learning~\cite{chen2014semi,gadde2014active}, social networks~\cite{perozzi2014deepwalk}, financial markets~\cite{saboksayr2021learning}, and traffic prediction~\cite{hasanzadeh2019piecewise}.
For an extensive overview of GSP, refer to~\cite{shuman2013emerging,sandryhaila2013discrete,ortega2018graph}.

Graph signal sampling is a key aspect of GSP and plays a crucial role in analyzing data on networks.
Many existing studies on graph signal sampling have focused on applying Shannon--Nyquist sampling theory to graph signals~\cite{chen2015discrete, marques2015sampling, anis2016efficient, tsitsvero2016signals, valsesia2018sampling, tanaka2018spectral, puy2018random, sakiyama2019eigendecomposition, bai2020fast, jayawant2022practical, wang2023fast}, including its extensions to time-varying graph signals~\cite{sheng2025subset}.
On the other hand, when representing real-world phenomena as graph signals, there are many cases that cannot be represented as bandlimited graph signals, such as meteorological data in mountainous areas, image data of landscapes, and data on water distribution networks~\cite{dong2016learning,cheung2018graph,li2025sensor}.

To accommodate such beyond bandlimited graph signals, extensions of the generalized sampling theory~\cite{elder2009beyond, eldar2015sampling} to the graph setting have been explored~\cite{tanaka2020sampling, tanaka2020gensamp}.
This theory provides a unified framework for sampling beyond bandlimited signals by assuming signal priors in the context of traditional signal sampling, and it aims to achieve not only perfect recovery but also the best possible recovery based on some strategies such as the least squares or minimax strategies.
By extending the theory to graph signals, it becomes possible to achieve sampling and recovery of beyond bandlimited graph signals.
However, in this extension, it is necessary to design a sampling operator, which was treated as a given operator when sampling traditional signals.

When extending the theory to the graph setting, there are two main approaches for sampling graph signals: sampling in the \textit{graph frequency domain} and sampling in the \textit{graph vertex domain}.
The authors of~\cite{chepuri2018graphsamp} pioneered the extension, which led to branches of sampling graph signals in the graph frequency domain~\cite{tanaka2020sampling,hara2023graph}.
Although graph frequency domain sampling methods are elegant, they are computationally expensive due to the need for graph Fourier transforms (GFT) and can be very time-consuming, especially for large graphs, so vertex domain methods have also been explored.

The approaches for sampling graph signals in the vertex domain~\cite{hara2021design,hara2022gsss,hara2024sensor} are to formulate an optimization problem to design a sampling operator that satisfies the condition regarding the recovery process to achieve the best possible recovery.
This condition can be reduced to a constraint that a matrix including the sampling operator has full rank.
However, handling this rank constraint is very difficult, especially in the framework of convex optimization.

\looseness=1 Vertex domain sampling methods can be classified into two primary strategies according to the design variables of the sampling operator: \textit{vertex-wise sampling} and \textit{aggregation sampling}.
In both strategies, the sampling operator can be expressed as the product of a selection matrix for a sampling set and an arbitrary linear graph filter, i.e., both of them belong to vertex domain sampling~\cite{tanaka2020sampling}.
Vertex-wise sampling determines the sampling set for a given graph filter.
Although each sampled value can be a linear combination of signal values across multiple vertices depending on the graph filter, the essence of vertex-wise sampling is the selection of the sampling set, which is a combinatorial problem.
In contrast, aggregation sampling directly designs the sampling operator, i.e., the weights for linearly combining signal values across multiple vertices, without predefining a graph filter.

\looseness=1 The method proposed in~\cite{hara2022gsss} adopts a vertex-wise sampling strategy and deals with arbitrary signal priors assumed in the generalized sampling theory.
This method has the difficulty of handling the constraint for achieving the best possible recovery in designing a sampling operator.
The constraint is handled as maximizing the determinant of a matrix including the sampling operator, though this is a combinatorial problem and is computationally difficult to solve.
Thus, as a compromise, it adopts a greedy algorithm that sequentially selects vertices one by one, which is computationally efficient and fast, especially for large graphs.
However, since it does not review the selected vertices during the optimization process once they are selected, it may include inappropriate vertices and fall into a low-quality suboptimal solution.
Also, the recovery accuracy is highly dependent on the selected vertices.
The accuracy may decrease if there is noise at the selected vertices, as the effect of the noise may be amplified in the recovery process.
Similarly, the method proposed in~\cite{hara2024sensor} employs a vertex-wise sampling strategy for sensor allocation.
This method also has the aforementioned weakness of vertex-wise sampling, and it relies on the strong assumption that the graph signal follows a specific signal prior.

Aggregation sampling approaches construct an \textit{aggregation sampling operator} that mixes signal values across multiple vertices to generate a sampled signal.
This leads to reduced dependence on specific vertices and allows for robust sampling and recovery by avoiding the effect of noise at the selected vertices to generate a sampled signal, which can be amplified in the recovery process.
To employ this strategy, the authors of~\cite{hara2021design} formulate a problem to design a sampling operator as an optimization problem with the constraint to achieve the best possible recovery based on the generalized sampling theory.
However, they apply an aggressive relaxation for handling the constraint for solving the optimization problem via convex optimization.
This relaxation may degrade the recovery accuracy.
Their method also relies on a strong assumption and can handle only a specific signal prior.

\looseness=1 Given the limitations of existing sampling methods, particularly those relying on a vertex-wise sampling strategy or convex optimization, a method for designing an aggregation sampling operator without applying strong assumptions or aggressive relaxations is desirable.
However, the rank constraint for achieving the best possible recovery is non-convex, which makes it difficult to handle in optimization problems.
Here, we focus on the fact that the rank constraint can be relaxed as a maximization of a nuclear norm, which is the tightest convex relaxation of the rank function on the unit ball of the spectral norm~\cite{fazel2002matrix}.
Then, it is possible to formulate the design of an aggregation sampling operator as a difference-of-convex (DC) optimization problem in the form of the difference between a function related to the structure of the sampling operator and the nuclear norm function.
A promising approach is to adopt a DC optimization framework to design an aggregation sampling operator without applying aggressive relaxations.
Here, a question arises:
\textit{How can we build a unified framework for designing an aggregation sampling operator applicable to beyond bandlimited graph signals under arbitrary signal priors via DC optimization for achieving the best possible recovery?}

In this paper, we propose a novel and unified framework for designing an aggregation sampling operator for beyond bandlimited graph signals assuming arbitrary priors in the vertex domain.
We formulate a design problem with the rank constraint to achieve the best possible recovery.
Then, for handling the rank constraint, we reformulate the problem as a DC optimization problem by introducing the nuclear norm relaxation of the constraint.
To solve the DC optimization problem, we develop a solver based on the general double-proximal gradient DC (GDPGDC) algorithm~\cite{banert2019general} introduced in Section~\ref{subsec:GDPGDC} with guaranteed convergence to a critical point.

The key contributions of this paper are summarized below:
\begin{itemize}[align=parleft, left=1.1mm, labelsep=1.1mm, itemindent=0mm, topsep=0pt, partopsep=0pt]
    \item Formulate the design of an aggregation sampling operator applicable to arbitrary signal priors based on the generalized sampling theory as an optimization problem with the rank constraint to achieve the best possible recovery.
    \item Transform the problem into a DC problem by relaxing the rank constraint using the nuclear norm.
    \item Develop a solver based on the GDPGDC algorithm for the DC problem with guaranteed convergence to a critical point.
\end{itemize}
We also conduct sampling and recovery experiments on various types of graph signals to validate the effectiveness.

Table~\ref{table:summary_existing_methods} summarizes the coverage of each method.
The column ``No GFT Required'' indicates whether the method avoids reliance on a graph Fourier transform.
The column ``Aggregation Sampling'' specifies whether the method adopts an aggregation sampling strategy.
The column ``Signal Priors'' indicates applicability to the signal priors.
\begin{table}[t]
    \captionsetup{font=footnotesize}
    \centering
    \caption{\footnotesize{Coverage of Existing and Proposed Methods.}}
    \vspace{-3mm}
    \label{table:summary_existing_methods}
    \renewcommand{\arraystretch}{1}
    \small
    \begin{tabular}[t]{c||c|c|c}
        \toprule
            Method & \shortstack{No GFT\\Required} & \shortstack{Aggregation\\Sampling} & \shortstack{Signal Priors} \\
        \midrule
            NLPD~\cite{chen2015discrete} & $\checkmark$ & - & Only Bandlimited \\ \hline
            AVM~\cite{jayawant2022practical} & $\checkmark$ & - & Only Bandlimited \\ \hline
            GSAO~\cite{wang2023fast} & $\checkmark$ & - & Only Bandlimited \\ \hline
            SASB~\cite{hara2021design} & $\checkmark$ & $\checkmark$ & Subspace \\ \hline
            GSSS~\cite{hara2022gsss} & $\checkmark$ & - & Arbitrary \\ \hline
            SUST~\cite{hara2023graph} & - & $\checkmark$ & Arbitrary \\ \hline
            $\textbf{Ours}$ & $\checkmark$ & $\checkmark$ & Arbitrary \\
        \bottomrule
    \end{tabular}
\end{table}

This paper is structured as follows.
Section~\ref{sec:preliminaries} reviews the generalized sampling theory and its extension to the graph setting.
The algorithm for solving the DC optimization problem is also introduced in this section.
In Section~\ref{sec:propose}, the proposed formulation using DC optimization and the developed solver are presented.
We demonstrate the effectiveness of the proposed method through experiments compared with existing methods by sampling and recovering graph signals in Section~\ref{sec:experiments}.
Finally, we conclude this paper in Section~\ref{sec:conclusion}.

The preliminary version of this work, which discusses a sampling method by assuming only a limited prior without proposing multiple designs of sampling operators, comprehensive experimental comparisons, and deeper discussions, has appeared in~\cite{yamashita2024apsipa}.

\section{Preliminaries}
\label{sec:preliminaries}
    \begin{table}[t]
    \captionsetup{font=footnotesize}
    \centering
    \caption{Notations and Definitions.}
    \vspace{-3mm}
    \label{table:notation}
    \renewcommand{\arraystretch}{1.3}
    \small
    \begin{tabular}[t]{c|c}
        \toprule
        Notations & Definitions \\
        \hline \hline
        $\set{X}$ & a set \\ \hline
        $\vect$, $\mat$ & a vector, a matrix \\ \hline
        $\vele, \left[\vect\right]_{i}$ & the $i$-th element of $\vect$  \\ \hline
        $\Ltwo{\vect}$ & the $\ell_2$ norm of $\vect$, $\Ltwo{\vect} := \sqrt{\sum_i \vele^2}$  \\ \hline
        $\ele, \left[\mat\right]_{ij}$ & the $(i,j)$-th element of $\mat$  \\ \hline
        $\tran{\mat}$ & the transpose of $\mat$  \\ \hline
        $\inv{\mat}$ & the inverse of $\mat$  \\ \hline
        $\pinv{\mat}$ & the pseudo-inverse of $\mat$  \\ \hline
        $\langle \mat, \matt \rangle$ & \begin{tabular}{c} the inner product of $\mat$ and $\matt$,\\$\langle \mat, \matt \rangle := \sum_{i}\sum_{j} \ele \elee$  \end{tabular}\\ \hline
        $\Lone{\mat}$ & the $\ell_1$ norm of $\mat$, $\Lone{\mat} := \sum_{i}\sum_{j} |\ele|$  \\ \hline
        $\FN{\mat}$ & the Frobenius norm of $\mat$, $\FN{\mat} := \sqrt{\langle \mat, \mat \rangle}$  \\ \hline
        $\NN{\mat}$ & the nuclear norm of $\mat$, $\NN{\mat} := \sum_{i} \sval{\mat}{i}$  \\ \hline
        $\sval{\mat}{i}$ & the $i$-th singular value of $\mat$  \\ \hline
        $\diag{(\cdot)}$ & \begin{tabular}{c}
            a diagonal matrix with $\cdot$\\as its principal diagonal
            \end{tabular}  \\ \hline
        $\indi{\set{C}}{\mat} $ & \begin{tabular}{c}
            the indicator function of a closed set $\set{C}$,\\ $\indi{\set{C}}{\mat} = \begin{cases}
                0 & \text{if } \mat \in \set{C} \\
                \infty & \text{otherwise}
            \end{cases}$ \\
            \end{tabular}  \\
        \bottomrule
    \end{tabular}
  \end{table}

    This section introduces mathematical tools required for the proposed method, in particular the generalized sampling theory and its extension to the graph setting.
    The fundamentals of GDPGDC, the algorithm needed to solve the proposed optimization problem, are also introduced.
    For reference, the notation is summarized in Table~\ref{table:notation}.

\subsection{Generalized Sampling} \label{sec:gs}

    In this subsection, we present the summary of the generalized sampling theory in Hilbert spaces~\cite{eldar2003sampling, eldar2015sampling}.
    Let $\sig$ be a vector in a Hilbert space $\HS$ and $\sampsig$ be its samples with the $i$-th sample given by
        $\elesampsig_i = \inprod{\sampRow_{(i)}}{\sig}$,
    where $\{\sampRow_{(i)}\}$ is a Riesz basis.
    The sampled signal $\sampsig$ is given by $\sampsig = \sampT \sig$, where $\samptranA$ is a sampling operator.
    From the sampled signal $\sampsig$, a recovered signal $\recsig$ is obtained by filtering $\sampsig$ with a correction operator $\cortran$ and a reconstruction operator $\rectran$ as follows:
    \begin{equation} \label{recsig}
        \recsig = \rectran \cortran \sampsig = \rectran \cortran \samptranA \sig,
    \end{equation}
    where $\rectran$ corresponds to a basis $\{\mathbf{w}_{i}\}$ for the reconstruction space, which spans a closed subspace $\set{W}$ of $\HS$.
    For achieving a stable recovery, $\{\mathbf{w}_{i}\}$ satisfies the Riesz basis condition~\cite{eldar2015sampling}.

    A reconstruction operator $\rectran$ may be constrained, i.e., it may be predefined for certain reasons, such as computational or hardware restrictions.
    We call the case where $\rectran$ is constrained the \textit{predefined case} and the case where $\rectran$ is not constrained the \textit{unconstrained case}.
    A correction operator $\cortran$ operates on the samples $\sampsig$ before the reconstruction.
    The major difference between the generalized sampling theory and the traditional sampling theory for bandlimited signals is the insertion of a correction operator $\cortran$ in the recovery process.
    By appropriately designing $\cortran$ using prior knowledge of the space to which the original signal $\sig$ belongs, it is possible to obtain a recovered signal that minimizes the error from the original signal in some sense.

\subsection{Framework of Graph Signal Sampling}
    \label{subsec:gs_sampling}
    In this subsection, we provide a brief overview of the generalized sampling theory in the graph setting~\cite{tanaka2020gensamp,tanaka2020sampling}, which forms the basis of our method.
    In this paper, we consider a weighted undirected graph $\graph = \{\gver, \gedge\}$, where $\gver$ and $\gedge$ denote a set of vertices and a set of edges between the vertices, respectively.
    The number of vertices is denoted as $N = |\gver|$.
    We define an adjacency matrix $\gadjM \in \realNN$, where each element $\eleadjM$ represents the weight of the edge between the $i$-th and $j$-th vertices.
    The weight from the $i$-th vertex to the $j$-th vertex and the weight from the $j$-th vertex to the $i$-th vertex are equal for undirected graphs, i.e., $\eleadjM = E_{ji}$.
    If the $i$-th and $j$-th vertices are unconnected, then $\eleadjM$ and $E_{ji}$ are set to $0$.
    The degree matrix $\gdegM \in \realNN$ is a diagonal matrix, where the $i$-th diagonal element $D_{ii}$ is defined as the sum of weights connected to the vertex $i$, i.e., $D_{ii} = \sum_{j}\eleadjM$.

    We use the graph Laplacian $\gLap := \gdegM - \gadjM$ as a graph variation operator for clarity and specificity.
    Since $\gLap$ is a real symmetric matrix, it always admits an eigendecomposition $\gLap = \guniM \geigM \guniMT$, where $\guniM = [\geigvec_1 \dots \geigvec_\numv]$ forms a unitary matrix containing the eigenvectors $\geigvec_{1}, \dots, \geigvec_{\numv}$, and $\geigM = \diag(\geigv_1,\dots,\geigv_\numv)$ comprises the eigenvalues $\geigv_i$.
    We refer to $\geigv_i$ as the $i$-th graph frequency, where $0 = \geigv_1 \leq \geigv_2 \leq \cdots \leq \geigv_\numv =: \geigv_{\max}$.

    Let $\gsig \in \gset \subseteq \realN$, $\gsampsig \in \realM (\nums \leq \numv)$, and $\grecsig \in \gset$ be an original graph signal, a sampled signal, and a recovered graph signal, respectively.
    A graph signal $\gsig$ undergoes sampling by a sampling operator $\sampMT \in \realMN$, i.e., $\gsampsig := \sampMT \gsig$.
    As described in Section~\ref{sec:intro}, both vertex-wise sampling and aggregation sampling belong to vertex domain sampling~\cite{tanaka2020sampling}, and they differ in the design variables.
    We define a sampling operator as follows:
    \begin{dfn}[Vertex-wise and aggregation sampling operators]
    \label{def:sampling_operators}
    \normalfont
    For a graph signal $\gsig \in \realN$ and a sampled signal $\gsampsig \in \realM$, a sampling operator $\sampMT \in \realMN$ determines how the samples are acquired.
    \begin{itemize}[align=parleft, left=1.1mm, labelsep=1.1mm, itemindent=0mm, topsep=0pt, partopsep=0pt]
        \item Vertex-wise sampling operator:
        As in~\cite[Definition~1]{tanaka2020sampling}, a vertex-wise sampling operator relies on selecting specific vertices and is defined by
        \begin{equation}
            \sampMT := \mathbf{I}_{\mathcal{M}} \mathbf{G},
            \label{eq:vw_sampling}
        \end{equation}
        where $\mathbf{I}_{\mathcal{M}} \in \{0, 1\}^{M \times N}$ is a binary selection matrix indicating the subset of sampled vertices $\mathcal{M} \subset \mathcal{V}$ ($|\mathcal{M}| = M$), and $\mathbf{G} \in \realNN$ is a given graph filter.
        The design variable of a vertex-wise sampling operator is the sampling set $\mathcal{M}$.
        When the graph filter $\mathbf{G}$ is not the identity matrix, each sampled value can be a linear combination of signal values across multiple vertices; however, the weights are fixed by the predefined $\mathbf{G}$.

        \item Aggregation sampling operator:
        In contrast, an aggregation sampling operator is not restricted by a combinatorial binary selection matrix or a predefined graph filter.
        It is defined as a general matrix $\sampMT \in \realMN$ whose entries themselves are the design variables, where each sampled value $c_{i}$ is obtained by a linear combination of the signal values across multiple vertices:
        \begin{equation}
            c_{i} = \sum_{j=1}^{N} [\sampMT]_{ij} x_{j}, \quad \text{for } i = 1, \dots, M.
            \label{eq:agg_sampling}
        \end{equation}
    \end{itemize}
    \end{dfn}
    The sampled signal $\gsampsig$ is filtered with a correction operator $\corM$ to reduce any errors or distortions introduced during the sampling and recovery process.
    Following this, it is further filtered by a reconstruction operator\footnote{
        We suppose $\recM^\top \recM$ is invertible under the predefined case for simplicity.
    } $\recM$ to map the sampled and corrected signal back onto the original graph.
    Hence, the recovered signal $\grecsig$ is represented as follows:
    \begin{equation}
        \label{gsig_recovery}
        \grecsig = \recM \corM \gsampsig = \recM \corM \sampMT \gsig.
    \end{equation}

    The recovery problem entails finding the optimal $\corM$ (and $\recM$ if it is unconstrained) based on assumed priors of the set $\gset$ to which $\gsig$ belongs.
    This framework encompasses various situations involving sampling and recovering, including scenarios where the graph signal is bandlimited (refer to~\cite{tanaka2020sampling} for more details).

    Based on the sampling and recovery framework described as Eq.~\eqref{gsig_recovery}, the correction operator $\corM$ and the reconstruction operator $\recM$ are designed by well-established strategies: least squares (LS) strategy, minimax (MX) strategy, and minimum mean squared error (MMSE) strategy.
    The LS strategy aims to find the recovered signal $\grecsig$ that minimizes the $\ell_2$ norm of the difference between the sampled signal created from the recovered signal, i.e., $\sampMT \grecsig$, and the sampled signal $\gsampsig$:
    \begin{equation} \label{LSstrategy}
        \grecsig_{\mathrm{LS}} = \argmin_{\grecsig \in \gset, \ \sampMT \gsig = \gsampsig} \Ltwo{ \sampMT \grecsig - \gsampsig}^{2}.
    \end{equation}
    The minimax strategy attempts to directly control the recovery error $\Ltwo{\grecsig - \gsig}^{2}$ for minimizing the error for the worst feasible signal:
    \begin{equation} \label{MXstrategy}
        \grecsig_{\mathrm{MX}} = \argmin_{\grecsig \in \gset} \max_{\gsig \in \gset, \ \sampMT \gsig = \gsampsig} \Ltwo{ \grecsig - \gsig}^{2}.
    \end{equation}
    The MMSE strategy aims to minimize the mean squared error (MSE) between $\grecsig$ and $\gsig$:
    \begin{equation} \label{MMSEstrategy}
        \grecsig_{\mathrm{MMSE}} = \argmin_{\grecsig \in \gset} \Expv{ \Ltwo{\grecsig - \gsig}^{2}},
    \end{equation}
    where $\Expv{\cdot}$ denotes the expected value.

    In graph signal sampling, the sampling operator $\sampMT$ needs to be designed carefully, since the assumptions on $\sampM$ differ between the generalized sampling theory for classical signal processing and graph signal sampling.
    While the generalized sampling theory typically assumes that $\sampM$ is given and determined by the acquisition device, in graph signal sampling, $\sampM$ must take into account graph structures and reflect dependencies between vertices, leading to a more complex design process for $\sampMT$.
    Thus, the sampling operator $\sampMT$ is designed by using $\corM$ and $\recM$ obtained from the strategies described above.

    This paper focuses on sampling and recovering graph signals under three representative signal priors assumed in the generalized sampling theory: subspace prior~\cite{tanaka2020sampling}, smoothness prior~\cite{tanaka2020sampling}, and stochastic prior~\cite{hara2023graph}.
    Next, we describe these priors and provide an overview of previous studies~\cite{tanaka2020gensamp,hara2023graph} related to the design of $\corM$ and $\recM$.
    The designs of the optimal $\corM$ and $\recM$ under these priors are summarized in Table~\ref{summary_corM_recM}.

    \begin{table*}[t]
    \centering
    \captionsetup{font=footnotesize}
    \caption{
        \footnotesize
        Designs of $\corM$ and $\recM$, Where $\smMXrecM = (\smopeMT\smopeM)^{-1}\sampM$ and $\smLSrecM = \recM (\recMT \smopeMT \smopeM \recM)^{-1} \recMT \sampM$.
    }
    \vspace{-3mm}
    \label{summary_corM_recM}
    \renewcommand{\arraystretch}{1.1}
    \small
    \begin{tabular}[t]{c|c||c|wc{2.5cm}|wc{2.5cm}}
        \toprule
            \multirow{2}{*}{Prior} & \multirow{2}{*}{Criteria} & Predefined & \multicolumn{2}{c}{Unconstrained} \\ \cline{3-5}
            &&$\corM$&$\corM$&$\recM$\\
        \hline
            \multirow{2}{*}{Subspace}
            & LS & $\pinv{(\sampMT \recM)}$ & \multirow{2}{*}{$\pinv{(\sampMT\sbgenM)}$} & \multirow{2}{*}{$\sbgenM$} \\
            & MX & $(\recMT\recM)^{-1}\recMT\sbgenM \pinv{(\sampMT\sbgenM)}$  & & \\
            \hline
            \multirow{2}{*}{Smoothness}
            & LS & $(\recMT \smopeMT \smopeM \recM)^{-1} \recMT \sampM \pinv{(\sampMT \smLSrecM)}$ & \multirow{2}{*}{$\pinv{(\sampMT \smMXrecM)}$} & \multirow{2}{*}{$\smMXrecM$}\\
            & MX & $(\recMT \recM)^{-1} \recMT \smMXrecM \pinv{(\sampMT \smMXrecM)}$ & & \\
            \hline
            Stochastic & MMSE & $(\recMT\recM)^{-1}\recMT\stsigM\sampM\pinv{(\sampMT\stsigM\sampM + \stnoiM)}$ & $\pinv{(\sampMT\stsigM\sampM + \stnoiM)}$ & $\stsigM\sampM$\\
        \bottomrule
    \end{tabular}
    \vspace{0mm}
\end{table*}

    \vspace{31bp}
    \subsubsection{Subspace Prior} \label{subsec:pre_sb_priors}
    We suppose that a graph signal $\gsig \in \realN$ is characterized by a linear model as follows:
    \begin{equation} \label{gs_sb}
        \gsig := \sbgenM \sbexpd,
    \end{equation}
    where $\sbgenM \in \realNK$ $(\sbgenn \leq \numv)$ is a known generator matrix and $\sbexpd \in \realK$ is an expansion coefficient vector.
    Here, we assume $\nums \geq \sbgenn$ for simplicity.
    This formulation encompasses the well-known bandlimited setting.
    Specifically, this refers to the case where the generator matrix $\sbgenM$ is a matrix with $\sbgenn$ graph Fourier components corresponding to $\sbgenn$ low graph frequencies, i.e., a matrix constructed by taking the first $\sbgenn$ columns of $\mathbf{U}$.

    For the unconstrained case, i.e., $\recM$ is not predefined, the solutions of the LS and minimax strategies are both given by
    \begin{equation}\label{sb_un_LS}
        \grecsig = \sbgenM \pinv{(\sampMT \sbgenM)} \sampMT \gsig,
    \end{equation}
    with the operators $\corM$ and $\recM$ given by
    \begin{equation}\label{sb_corM_recM}
        \corM = \pinv{(\sampMT \sbgenM)}, \quad \recM = \sbgenM.
    \end{equation}

    For the predefined case, i.e., $\recM$ is predefined, we take the LS and minimax strategies for designing the correction operator $\corM$.
    The solution of the LS strategy is
    \begin{equation}\label{sb_sol_LS}
        \grecsig = \recM \pinv{(\sampMT \recM)} \sampMT \gsig,
    \end{equation}
    where the correction operator $\corM$ is
    \begin{equation}\label{sb_corM_LS}
        \corM = \pinv{(\sampMT \recM)},
    \end{equation}
    and the solution of the minimax strategy is
    \begin{equation}\label{sb_sol_MX}
        \grecsig = \recM (\recMT\recM)^{-1}\recMT\sbgenM \pinv{(\sampMT\sbgenM)} \sampMT \gsig,
    \end{equation}
    where the correction operator $\corM$ is
    \begin{equation}\label{sb_corM_MX}
        \corM = (\recMT\recM)^{-1}\recMT\sbgenM\pinv{(\sampMT\sbgenM)}.
    \end{equation}

    \vspace{5bp}
    \subsubsection{Smoothness Prior} \label{subsec:pre_sm_priors}
    The smoothness prior is less restrictive than the subspace prior in that the signal subspace is not assumed to be known.
    Specifically, we assume that the signal $\gsig$ is smooth in the following sense: $\Ltwo{\smopeM \gsig}^{2} \leq \smbound^2$ for some constant $\smbound$, where $\smopeM \in \realNN$ is an invertible operator that quantifies the variation of $\gsig$.
    A typical example of $\smopeM$ is a graph filter $\smopeM = \guniM F(\geigM) \guniMT$, where $F(\geigM) = \diag(F(\geigv_1), \dots, F(\geigv_\numv))$ and $F(\geigv)$ is a spectral response increasing with the graph frequency; for example, $F(\geigv) = \geigv/\geigv_{\max} + \delta$ with $\delta > 0$, i.e., $\smopeM = \gLap/\geigv_{\max} + \delta \mathbf{I}$.
    In this case, $\Ltwo{\smopeM \gsig}$ is small when $\gsig$ is dominated by low graph frequency components, i.e., when $\gsig$ is smooth on the graph.
    Here we consider two approaches to select a solution, the LS and minimax strategies, which can be applied in both the unconstrained and predefined cases.

    For the unconstrained case, the solution of the LS strategy is given by
    \begin{equation}\label{sm_un_sol_LS}
        \grecsig = \smMXrecM \pinv{(\sampMT\smMXrecM)} \sampMT \gsig,
    \end{equation}
    where $\smMXrecM = (\smopeMT\smopeM)^{-1}\sampM$, with the following $\corM$ and $\recM$:
    \begin{equation} \label{sm_un_corM_recM}
        \corM = \pinv{(\sampMT \smMXrecM)}, \quad
        \recM = \smMXrecM = \inv{(\smopeMT\smopeM)}\sampM.
    \end{equation}
    The solution of the minimax strategy coincides with Eq.~\eqref{sm_un_sol_LS}.

    For the predefined case, the solution of the LS strategy is given by
    \begin{equation}\label{sm_pre_sol_LS}
    \grecsig = \smLSrecM \pinv{(\sampMT \smLSrecM)} \sampMT \gsig,
    \end{equation}
    where $\smLSrecM = \recM \inv{(\recMT \smopeMT \smopeM \recM)} \recMT \sampM$.
    Thus, the correction operator $\corM$ is
    \begin{equation} \label{sm_pre_LS_corM}
    \corM = (\recMT \smopeMT \smopeM \recM)^{-1} \recMT \sampM \pinv{(\sampMT \smLSrecM)}.
    \end{equation}
    The solution of the minimax strategy is given by
    \begin{equation}\label{sm_pre_sol_MX}
        \grecsig = \recM (\recMT\recM)^{-1}\recMT \smMXrecM \pinv{(\sampMT \smMXrecM)} \sampMT \gsig,
    \end{equation}
    with the following $\corM$:
    \begin{equation} \label{sm_pre_MX_corM}
    \corM = (\recMT \recM)^{-1} \recMT \smMXrecM \pinv{(\sampMT \smMXrecM)}.
    \end{equation}

    \vspace{5.2bp}
    \subsubsection{Stochastic Prior} \label{subsec:pre_st_priors}
    We consider that the samples can be obtained with additive noise $\stnoise \in \realM$ and recover the signal from the noisy sampled signal $\ssign = \gsampsig + \stnoise$ as:
    \begin{equation} \label{st-reconstruction}
        \grecsig = \recM \corM \ssign = \recM \corM (\gsampsig + \stnoise) = \recM \corM (\sampMT \gsig + \stnoise).
    \end{equation}

    Suppose that a graph signal $\gsig$ is a zero-mean process characterized by graph wide sense stationarity (GWSS)~\cite{hara2023graph}, which parallels the wide sense stationarity (WSS) typically observed in standard signals.
    The graph signal $\gsig$ is associated with a known covariance matrix $\stsigM \in \realNN$.
    Since $\gsig$ is GWSS, $\stsigM$ is diagonalized by the graph Fourier basis, i.e., $\stsigM = \guniM \widehat{\Gamma}_{\gsig}(\geigM) \guniMT$, where $\widehat{\Gamma}_{\gsig}(\geigM) = \diag(\widehat{\Gamma}_{\gsig}(\geigv_1), \dots, \widehat{\Gamma}_{\gsig}(\geigv_\numv))$ is the graph power spectral density~\cite{hara2023graph, perraudin2017stationary}.
    For example, $\widehat{\Gamma}_{\gsig}(\geigv) = 1/(\geigv + \delta)$ with $\delta > 0$, i.e., $\stsigM = (\gLap + \delta \mathbf{I})^{-1}$, represents a graph signal whose power is concentrated on low graph frequencies.

    Additionally, let $\stnoise$ be another zero-mean random vector, possessing a known covariance matrix $\stnoiM \in \realMM$.
    It is important to note that these processes, $\gsig$ and $\stnoise$, are independent of each other.
    Both $\stsigM$ and $\stnoiM$ are autocorrelation matrices in this context.

    For the unconstrained case, as $\recM$ can be freely chosen along with $\corM$, the solution of the MMSE strategy is given by
    \begin{equation}\label{st_un_MMSE}
        \grecsig = \stsigM \sampM \pinv{(\sampMT \stsigM \sampM + \stnoiM)} (\sampMT \gsig + \stnoise),
    \end{equation}
    with the operators $\corM$ and $\recM$ given by:
    \begin{equation} \label{st_un_corM_recM}
        \corM = \pinv{(\sampMT \stsigM \sampM + \stnoiM)}, \quad
        \recM = \stsigM \sampM.
    \end{equation}

    For the predefined case, the solution of the MMSE strategy is given by
    \begin{equation}\label{st_pre_MMSE}
        \scalebox{0.95}{$\grecsig = \recM (\recMT \recM)^{-1} \recMT \stsigM \sampM \pinv{(\sampMT \stsigM \sampM + \stnoiM)} (\sampMT \gsig + \stnoise),$}
    \end{equation}
    where the correction operator $\corM$ is
    \begin{equation}\label{st_pre_corM}
        \corM = (\recMT \recM)^{-1} \recMT \stsigM \sampM \pinv{(\sampMT \stsigM \sampM + \stnoiM)}.
    \end{equation}

\PublicationFlushDisplayColumn{2.851bp}
\setcounter{topnumber}{2}
\subsection{GDPGDC Algorithm}
\label{subsec:GDPGDC}
\begin{publicationdisplaypadding}{4.5pt}
    Having discussed the recovery of graph signals based on the generalized sampling theory, we now focus on an algorithm for solving DC minimization problems.
    In this subsection, we introduce the GDPGDC algorithm~\cite{banert2019general}, a versatile tool for solving DC minimization problems.

    The GDPGDC algorithm can solve problems of the form
    \begin{equation}
        \label{prob:DC_general}
        \min_{\mat} f_1(\mat) + f_2(\mat) - h(\mattt) \quad \st \ \mattt = \matDC \mat,
    \end{equation}
    where $f_1: \realnn{n}{m} \rightarrow \realnum$ is a differentiable convex function with a $1/\beta$-Lipschitz continuous gradient for some $\beta > 0$, $f_2:\realnn{n}{m} \rightarrow \exR$ and $h:\realnn{k}{m} \rightarrow \exR$ are proper lower-semicontinuous convex functions for $\exR := \realnum \cup \{\infty\}$, and $\matDC \in \realnn{k}{n}$ is a matrix.

    Then, the GDPGDC algorithm solves Prob.~\eqref{prob:DC_general} by the following procedures: for $\gamma_1 > 0$ and $\gamma_2 > 0$, iterate
    \begin{equation}
        \label{eq:DCAbyGDPG}
        \resizebox{\linewidth-0mm}{!}{$
        \left\lfloor
        \begin{array}{l}
            \mat^{(\Step + 1)} \leftarrow \prox_{\gamma_{1} f_{2}} \left(\mat^{(\Step)} - \gamma_{1} \left(\nabla f_{1} \left(\mat^{(\Step)}\right) - \tran{\matDC} \mathbf{Z}^{(\Step)}\right)\right); \\
            \mattt^{(\Step+1)} \leftarrow \prox_{\gamma_{2}h^{*}} \left(\mattt^{(\Step)} + \gamma_{2}\matDC \mat^{(\Step + 1)}\right);\\
            \Step \leftarrow \Step + 1;
        \end{array}
        \right.
        $}
    \end{equation}

    Here, the proximity operator of a proper lower-\hspace{0pt}semicontinuous convex function $f$ with a parameter $\gamma > 0$ is defined as~\cite{bauschke2017correction}
    \begin{equation} \label{eq:prox_f2}
    \begin{aligned}
        &\prox_{\gamma f}:\realnn{n}{m}\rightarrow\realnn{n}{m}: \\
        &\matt \mapsto \argmin_{\mat} f(\mat) + \frac{1}{2\gamma} \FN{\matt - \mat}^{2}.
    \end{aligned}
    \end{equation}

    The \textit{Fenchel--Rockafellar conjugate function} of $f$ is represented as $f^{*}$ and defined as
    \begin{equation}
        f^*(\mat):=\sup_{\matt}\langle \mat, \matt \rangle - f(\matt).
    \end{equation}
    The proximity operator of $f^*$ is calculated with a parameter $\gamma > 0$ thanks to Moreau's identity~\cite[Theorem~14.3(ii)]{bauschke2017correction} as
    \begin{equation} \label{eqDCh}
        \prox_{\gamma f^*}(\mat) = \mat - \gamma \prox_{\frac{1}{\gamma} f}\left(\frac{1}{\gamma}\mat\right).
    \end{equation}

\end{publicationdisplaypadding}
    We summarize the theoretical results for the convergence of the GDPGDC algorithm as follows:
    \begin{theo}
    [{Convergence of the sequence generated by the GDPGDC algorithm~\cite[Proposition~4 and Theorem~1]{banert2019general}}]
    \label{theo:convergence_GDPGDC}
    Let $\Phi(\mat, \mattt) := f_1(\mat) + f_2(\mat) + h^{*}(\mattt) - \langle \mattt, \matDC \mat \rangle$, let $\inf \{ f_1(\mat) + f_2(\mat) - h(\matDC \mat) \; | \; \mat \in \realnn{n}{m} \} > - \infty$, and let $0 < \gamma_1 < 2 \beta$ and $0 < \gamma_2 < +\infty$ be satisfied.
    If $\{\mat^{(\Step)}\}_{\Step \in \mathbb{N}}$ and $\{\mattt^{(\Step)}\}_{\Step \in \mathbb{N}}$, generated by~\eqref{eq:DCAbyGDPG}, are bounded, then every cluster point of $\{\mat^{(\Step)}\}_{\Step \in \mathbb{N}}$ is a critical point\footnote{
        The definition of a critical point is as follows~\cite[Definition~1]{banert2019general}:
        A point $(\overline{\mathbf{X}}, \overline{\mathbf{Z}})$ is defined as a critical point of $\Phi$ if it satisfies $\mathbf{B}^{\top}\overline{\mathbf{Z}} \in \nabla f_1(\overline{\mathbf{X}}) + \partial f_2(\overline{\mathbf{X}})$ and $\mathbf{B}\overline{\mathbf{X}} \in \partial h^{*}(\overline{\mathbf{Z}})$, where $\partial$ denotes the convex subdifferential.
        In this case, $\overline{\mathbf{X}}$ is called a critical point of Prob.~\eqref{prob:DC_general}.
    }
    of Prob.~\eqref{prob:DC_general}.
    In addition, if $0 < \gamma_1 < \beta$ and $\Phi$ is a Kurdyka--{\L}ojasiewicz (K{\L}) function~\cite[Definition~2]{banert2019general}, then $\{(\mat^{(\Step)}, \mattt^{(\Step)})\}_{\Step \in \mathbb{N}}$ converges to a critical point of $\Phi$, and thus $\{\mat^{(\Step)}\}_{\Step \in \mathbb{N}}$ converges to a critical point of Prob.~\eqref{prob:DC_general}.
    \end{theo}

\newpage
\section{Proposed Method}
\label{sec:propose}
\looseness=-1 The main challenge in graph signal sampling based on the generalized sampling theory is designing the sampling operator $\sampMT$.
In this section, we present our proposed method for its design.
We formulate the design of an aggregation sampling operator as a DC optimization problem.
Our formulation provides a unifying framework applicable to any priors assumed in the generalized sampling theory.
Finally, we develop a solver based on the GDPGDC algorithm to solve the problem.

\vspace{-8.7bp}
\subsection{Conditions for Designing the Sampling Operator}
\label{subsec:condition_for_sampling_operator}
\looseness=-1 In this subsection, we discuss the conditions for designing a sampling operator for each prior, using the solutions for the recovered signals introduced in Section~\ref{subsec:gs_sampling} before discussing the specific design method for the sampling operator.
In order to achieve the best possible recovery based on Eqs.~\eqref{gsig_recovery} and~\eqref{st-reconstruction}, we try to find $\sampM$ ensuring that the correction operator $\corM$, which involves a pseudo-inverse operation as summarized in Table~\ref{summary_corM_recM}, has full rank for the recovery process to be numerically stable and for the resulting optimal solution to be unique.
We will now examine the conditions on $\sampM$ on a prior-by-prior basis to ensure the desired properties.

\subsubsection{Subspace Prior}
Based on Eqs.~\eqref{sb_un_LS} and~\eqref{sb_sol_MX}, we seek $\sampM$ for which $\pinv{(\sampMT\sbgenM)}$ has full row rank, i.e., $\sampMT\sbgenM$ has full column rank for the unconstrained case and under the minimax strategy for the predefined case.
Thus, we need to find $\sampM$ such that $\sbgenMT \sampM$ has full row rank for these cases.

\looseness=-1 Similarly, based on Eq.~\eqref{sb_sol_LS}, we seek $\sampM$ for which $\pinv{(\sampMT\recM)}$ has full rank under the LS strategy for the predefined case.
Thus, we need to find $\sampM$ such that $\recMT \sampM$ has full row rank for this case.\footnote{
    Since $\recM \in \realNM$, $\recMT \sampM$ is an $\nums \times \nums$ square matrix, and thus its full row rank is equivalent to its invertibility.
}

\subsubsection{Smoothness Prior}
Based on Eqs.~\eqref{sm_un_sol_LS} and~\eqref{sm_pre_sol_MX}, for the unconstrained case and under the minimax strategy for the predefined case, we seek $\sampM$ for which $\pinv{(\sampMT \smMXrecM)} = \inv{(\sampMT \smMXrecM)}$, i.e., $\sampMT \smMXrecM$ is invertible.
Let us define the singular value decomposition (SVD) of $\smopeM$ as $\smopeM = \smsvul \smsv \smsvur^\top$.
Then, $\sampMT \smMXrecM$ is transformed as follows:
\begin{align}
    \sampMT \smMXrecM &= \sampMT (\smopeMT\smopeM)^{-1}\sampM
    = \sampMT (\smsvur \smsv^{2} \smsvur^\top)^{-1}\sampM \nonumber \\
    &= (\smsv^{-1} \smsvur^\top \sampM)^\top \smsv^{-1} \smsvur^\top \sampM.
\end{align}
Thus, we need to find $\sampM$ such that $\smsv^{-1} \smsvur^\top \sampM \in \realNM$ has full column rank for these cases.

Similarly, based on Eq.~\eqref{sm_pre_sol_LS}, we seek $\sampM$ for which $\pinv{(\sampMT \smLSrecM)} \allowbreak = \inv{(\sampMT \smLSrecM)}$, i.e., $\sampMT \smLSrecM$ is invertible under the LS strategy for the predefined case.
Let us define the economy-size SVD of $\smopeM \recM \in \realNM$ as $\smopeM \recM = \smsvorul \smsvor \smsvorur^{\top}$, where $\smsvorul \in \realNM$ and $\smsvorur^{\top} \in \realMM$ are the matrix of left singular vectors and the transpose of the matrix of right singular vectors, respectively, and $\smsvor \in \realMM$ is the singular values matrix, which contains only non-zero singular values.
Then, $\sampMT \smLSrecM$ is transformed as follows:
\begin{align}
    \sampMT \smLSrecM &= \sampMT \recM (\recMT \smopeMT \smopeM \recM)^{-1} \recMT \sampM \nonumber \\
    &= \sampMT \recM \smsvorur \smsvor^{-2} \smsvorur^{\top} \recMT \sampM \nonumber \\
    &= (\smsvor^{-1} \smsvorur^{\top} \recMT \sampM)^{\top} (\smsvor^{-1} \smsvorur^{\top} \recMT \sampM).
\end{align}
Thus, we need to find $\sampM$ such that $\smsvor^{-1} \smsvorur^{\top} \recMT \sampM \in \realMM$ has full column rank for this case.

\subsubsection{Stochastic Prior}
Based on Eqs.~\eqref{st_un_MMSE} and~\eqref{st_pre_MMSE}, for both the unconstrained and predefined cases, we seek $\sampM$ for which $\pinv{(\sampMT\stsigM\sampM + \stnoiM)} = \inv{(\sampMT\stsigM\sampM + \stnoiM)}$, i.e., $\sampMT\stsigM\sampM + \stnoiM$ is invertible.
Due to the characteristics of an autocorrelation matrix, $\sampMT\stsigM\sampM + \stnoiM$ is invertible if $\sampMT\stsigM\sampM$ is invertible.
Since $\stsigM$ is symmetric positive semidefinite, for some $\stsqM \in \realNN$, $\stsigM$ can be decomposed as $\stsigM = \stsqM^\top \stsqM$.
Then $\sampMT\stsigM\sampM = (\stsqM \sampM)^\top \stsqM \sampM$ is invertible if and only if $\stsqM \sampM \in \realNM$ has full column rank for both the unconstrained and predefined cases.

Based on the above discussion, we introduce a matrix $\varM$ to find an appropriate $\sampM$ that satisfies the condition that $\corM$ has full rank for each prior.
Here, the condition for achieving the best possible recovery is unified as the condition that $\varM \sampM$ has full rank.
The components of the matrix $\varM$ for each prior are summarized in Table~\ref{table:P}.

\begin{table}[t]
    \captionsetup{font=footnotesize}
    \caption{\footnotesize{
        Components of the Matrix $\varM$, where
            $\smopeM = \smsvul \smsv \smsvur^\top$,
            $\smopeM \recM = \smsvorul \smsvor \smsvorur^{\top}$, and
            $\stsigM = \stsqM^\top \stsqM$.
    }}
    \vspace{-3mm}
    \label{table:P}
    \renewcommand{\arraystretch}{1.3}
    \small
    \begin{tabular}[t]{c|c||c|c}
        \toprule
            Prior & Criteria & Predefined & Unconstrained\\
            \hline
            \multirow{2}{*}{Subspace} & LS & $\recMT$ & \multirow{2}{*}{$\sbgenM^{\top}$} \\
            & MX & $\sbgenM^{\top}$ & \\
            \hline
            \multirow{2}{*}{Smoothness} & LS & $\smsvor^{-1} \smsvorur^{\top} \recMT$ & \multirow{2}{*}{$\smsv^{-1} \smsvur^{\top}$}\\
            & MX & $\smsv^{-1} \smsvur^{\top}$ &  \\
            \hline
            Stochastic & MMSE & $\stsqM$ & $\stsqM$ \\
        \bottomrule
    \end{tabular}
\end{table}

\vspace{10.8bp}
\subsection{Problem Formulation}
\label{subsec:prob_formulation}
In this subsection, we discuss the specific method of designing a sampling operator $\sampMT$.
Assuming that we have defined $\recM$ and $\corM$ as summarized in Table~\ref{summary_corM_recM}, our current task is to devise an appropriate $\sampM$ that aligns with the recovery process described in Eqs.~\eqref{gsig_recovery} and~\eqref{st-reconstruction}.

Now, the question arises as to how to design such $\sampM$.
To address this, we first formulate the sampling operator design problem as follows:
\begin{align}\label{prob:org}
    \min_{\sampM}\fsampM \ \st \ \sampM \in \sampMSet \mbox{ and } \varM \sampM \mbox{ has full rank},
\end{align}
\looseness=-1 where $\fsampM$ is a proper lower-semicontinuous convex function that controls the structure of $\sampM$ such as sparsity or smoothness of its elements, and $\sampMSet$ is a nonempty closed convex set.
The first constraint serves to control the elements of the sampling operator for ensuring stability within the set $\sampMSet$.
The second constraint is to achieve the best possible recovery; however, this constraint is difficult to handle directly.
Since the guarantee of full rankness of $\varM \sampM$ indicates that $-\rank(\varM \sampM)$ should be as small as possible, we rewrite Prob.~\eqref{prob:org} as follows:
\begin{align}\label{prob:org3}
    \min_{\sampM} \fsampM - \rank({\varM \sampM}) \ \st \ \sampM\in \sampMSet.
\end{align}
Prob.~\eqref{prob:org3} is still difficult to optimize due to the combinatorial nature of the rank function.
Therefore, by introducing the nuclear norm, which is known as the tightest convex relaxation of the rank function on the unit ball of the spectral norm~\cite{fazel2002matrix}, we relax the problem as
\begin{align}\label{prob:org4}
    \min_{\sampM} \fsampM - \NN{\varM \sampM} \ \st \ \sampM\in \sampMSet.
\end{align}

We consider three specific designs for $\fsampM$ and $\sampMSet$:
\begin{itemize}[align=parleft, left=1.1mm, labelsep=1.1mm, itemindent=0mm, topsep=0pt, partopsep=0pt]
    \item Design (i): Limit the Frobenius norm of $\sampM$ to at most $\paramLBall >0$, i.e.,
    \begin{align} \label{prob:Lball}
        \fsampM = 0, \
        \sampMSet = \LBall := \{ \mat \; | \; \FN{\mat} \leq \paramLBall \}.
    \end{align}
    This allows us to control the magnitude of $\sampM$ and prevent the excessive amplification of the graph signal.
    \item Design (ii): Minimize the square of the Frobenius norm of $\sampM$ while its elements are limited between $\lbound$ and $\ubound$, i.e.,
    \begin{align} \label{prob:FNandBox}
        \fsampM = \paramBox \FN{\sampM}^2, \
        \sampMSet = \BoxconNM,
    \end{align}
    where $\paramBox$ is a balancing parameter.
    This design encourages a solution where the magnitude is spread out evenly across all elements of $\sampM$ while its elements are constrained to be within the range $[\lbound, \ubound]$.
    In other words, this reduces large values and increases small values, which leads to a solution having small fluctuations.
    \item Design (iii): Minimize the $\ell_1$ norm of $\sampM$ while its elements are limited between $\lbound$ and $\ubound$, i.e.,
    \begin{align} \label{prob:L1andBox}
        \fsampM = \paramBox \Lone{\sampM}, \
        \sampMSet = \BoxconNM,
    \end{align}
    where $\paramBox$ is a balancing parameter.
    This design tends to promote sparsity by minimizing the sum of the absolute values of the elements of $\sampM$, i.e., it encourages $\sampM$ to have many zero or near-zero elements, while its elements are constrained to the range $[\lbound, \ubound]$.
    This approach allows focused sampling on specific vertices, thereby facilitating the effective capture of crucial portions of the graph.
\end{itemize}

\begin{rem}[Trends of sampling operators designed by the proposed method]
    \rm{
        As an aggregation sampling operator tends to use values across the entire graph, it requires communication between vertices in distributed implementations.
        Thus, there may be a concern that the proposed method is not suitable for applications that require sampling only a few vertices such as sensor placement problems.
        However, by imposing specific constraints on $\sampMT$, it would be possible to achieve local observations; for instance, restrict some columns of $\sampMT$ to be zero vectors so that the corresponding vertices are not sampled by the operator.
        Design (iii) addresses this by promoting sparsity in $\sampM$ via $\ell_1$-norm minimization, acting as local observations to mitigate communication overhead while maintaining the advantages of aggregation sampling.
        Furthermore, to address such concerns, methods for designing an aggregation sampling operator while controlling the number of active vertices have been proposed in~\cite{yamashita2025controlling, yamashita2026sampling}.
    }
\end{rem}

\begin{algorithm}[t]
    \caption{Sampling operator design algorithm}
    \label{algo:GDPG}
    \begin{algorithmic}[1]
    \renewcommand{\algorithmicrequire}{\textbf{Input:}}
    \renewcommand{\algorithmicensure}{\textbf{Output:}}
    \REQUIRE $\varM, \fsamp, \sampMSet, \sampM^{(0)}, \DCw^{(0)}, \gamma_1>0, \gamma_2>0$
    \WHILE{a stopping criterion is not satisfied}
    \STATE $ \sampM^{(\Step+1)} \leftarrow \prox_{\gamma_1(\fsamp + \iota_{\sampMSet})}\left( \sampM^{(\Step)} + \gamma_{1} \varM^\top \DCw^{(\Step)}\right)$
    \\by~\eqref{eq:proj_Fball},~\eqref{eq:proj_FNandBox}, or~\eqref{eq:proj_L1andBox}; \label{al_s}
    \STATE $ \DCw^{(\Step+1)} \leftarrow \prox_{\gamma_{2} (\NN{\cdot})^{*}} \left(\DCw^{(\Step)} + \gamma_{2}\varM \sampM^{(\Step + 1)}\right)$
    \\by~\eqref{eqDCh} and~\eqref{eq:prox_NN1}; \label{al_z}
    \STATE $ \Step \leftarrow \Step+1 $;
    \ENDWHILE
    \ENSURE $\sampM^{(\Step)}$
    \end{algorithmic}
    \vspace{10bp}
\end{algorithm}

\begin{rem}[Full rankness of $\varM \sampM$]
\rm{
    Prob.~\eqref{prob:org4} utilizes the nuclear norm as a convex relaxation for the rank function to ensure the problem is tractable.
    While this relaxation does not theoretically guarantee that the designed sampling operator $\sampM$ will always make $\varM \sampM$ have full rank, our experimental evaluations across all cases show that $\varM \sampM$ consistently has full rank.
    While a random matrix with independent and identically distributed (i.i.d.) entries might satisfy the full rank condition, our framework prioritizes numerical stability by minimizing $-\NN{\varM \sampM}$ to push the singular values of $\varM \sampM$ away from zero.
    This optimization improves the operator's condition number, thereby enhancing robustness against noise and approximation errors compared to a simple random initialization.
}
\end{rem}

\subsection{Optimization}
In this subsection, we describe the optimization algorithm for solving Prob.~\eqref{prob:org4}.
By introducing the indicator function of $\sampMSet$, i.e., $\iota_{\sampMSet}$, which is also a proper lower-semicontinuous convex function, we can reformulate Prob.~\eqref{prob:org4} as follows:
\begin{align}\label{prob:proposed}
    \min_{\sampM} \fsampM + \iota_{\sampMSet}(\sampM) - \NN{\varM \sampM}.
\end{align}

This problem is regarded as the minimization of the difference between two convex functions, and Prob.~\eqref{prob:proposed} is reduced to Prob.~\eqref{prob:DC_general} since $\fsampM + \iota_{\sampMSet}(\sampM)$ and $\NN{\varM \sampM}$ are both proper lower-semicontinuous convex functions.
Thus, the GDPGDC algorithm can be applied to solve this problem.

The algorithmic procedure for solving the problem is summarized in Algorithm~\ref{algo:GDPG} with the respective functions and variables in Prob.~\eqref{prob:DC_general} as follows: $f_1(\mat) = 0$, $f_2(\mat) = \fsamp(\mat) + \indi{\sampMSet}{\mat}$, $h(\mattt) = \NN{\mattt}$, $\mat = \sampM$, and $\matDC = \varM$.

In what follows, we derive specific computations of each step of the algorithm.
For Step~\ref{al_s} of Algorithm~\ref{algo:GDPG}, as there are three specific designs for $\fsamp$ and $\sampMSet$, there are also three patterns of the proximity operator $\prox_{\gamma_1(\fsamp + \iota_{\sampMSet})}(\mat)$ as follows:
\begin{itemize}[align=parleft, left=1.1mm, labelsep=1.1mm, itemindent=0mm, topsep=0pt, partopsep=0pt]
    \item Design (i): Since $\fsampM = 0$ and $\sampMSet = \LBall$ is a nonempty closed convex set, the proximity operator of $\gamma_1(\fsamp + \iota_{\sampMSet})$ is equal to the metric projection\footnote{The proximity operator of the indicator function of a nonempty closed convex set $\set{C}$ is identical to the metric projection onto $\set{C}$~\cite{bauschke2017correction}.} onto $\LBall$, i.e.,
    \begin{align}
        \label{eq:proj_Fball}
        \prox_{\gamma_1(\fsamp + \iota_{\sampMSet})}(\mat)
        &= \prox_{\iota_{\LBall}} (\mat) \nonumber \\
        &= \begin{cases}
            \mat, & \mathrm{if} \: \mat \in \LBall; \\
            \frac{\paramLBall \mat}{\FN{\mat}}, & \mathrm{otherwise}.
        \end{cases}
    \end{align}
    \item Design (ii): As $\fsampM = \paramBox \FN{\sampM}^2$ and $\sampMSet = \BoxconNM$, the proximity operator $\prox_{\gamma_1(\fsamp + \iota_{\sampMSet})}(\mat)$ is calculated as follows: for all $i$ from $1$ to $\numv$ and all $j$ from $1$ to $\nums$,
    \begin{align}
        \label{eq:proj_FNandBox}
        &\left[\prox_{\gamma_1(\fsamp + \iota_{\sampMSet})}(\mat)\right]_{ij} \nonumber \\
        &= \left[\prox_{\gamma_1( \paramBox \FN{\cdot}^2 + \indi{\BoxconNM}{\cdot})} (\mat)\right]_{ij} \nonumber \\
        &= \max \left\{ \lbound, \min \left\{ \prox_{\gamma_1 \paramBox \FN{\cdot}^2} (\ele), \ubound \right\}\right\}.
    \end{align}
    For the details of this calculation, see Appendix~\ref{appendix:calc_FNandBox}.
    \item Design (iii): As $\fsampM = \paramBox \Lone{\sampM}$ and $\sampMSet = \BoxconNM$, the proximity operator $\prox_{\gamma_1(\fsamp + \iota_{\sampMSet})}(\mat)$ is calculated as follows: for all $i$ from $1$ to $\numv$ and all $j$ from $1$ to $\nums$,
    \begin{align}
        \label{eq:proj_L1andBox}
        &\left[\prox_{\gamma_1(\fsamp + \iota_{\sampMSet})}(\mat)\right]_{ij} \nonumber \\
        &= \left[\prox_{\gamma_1( \paramBox \Lone{\cdot} + \indi{\BoxconNM}{\cdot})} (\mat)\right]_{ij} \nonumber \\
        &= \max \left\{ \lbound, \min\left\{ \prox_{\gamma_1 \paramBox \Lone{\cdot}}\left(\ele\right), \ubound \right\}\right\}.
    \end{align}
    For the details of this calculation, see Appendix~\ref{appendix:calc_L1andBox}.
\end{itemize}

Next, as for Step~\ref{al_z} of Algorithm~\ref{algo:GDPG}, we present the specific computation of the proximity operator of the nuclear norm.
The proximity operator $\prox_{\gamma\NN{\cdot}}(\mat)$ is calculated by
\begin{equation}
  \label{eq:prox_NN1}
  \prox_{\gamma\NN{\cdot}}(\mat)=\mathbf{U}_\mat \mathcal{S}_{\gamma}(\mathbf{\Sigma}_{\mat}) \mathbf{V}_\mat^\top,
\end{equation}
where $\mat = \mathbf{U}_{\mat} \mathbf{\Sigma}_{\mat} \mathbf{V}_{\mat}^\top$ is the SVD of $\mat$, and $\mathcal{S}_{\gamma}(\cdot)$ is the soft-thresholding operator applied to singular values as:
\begin{equation}
  \mathcal{S}_{\gamma}(\sigma_i) = \max(\sigma_i - \gamma, 0),
\end{equation}
and $\mathcal{S}_{\gamma}(\mathbf{\Sigma}_{\mat})$ is a diagonal matrix with entries $\mathcal{S}_{\gamma}(\sval{\mat}{i})$, i.e., $\mathcal{S}_{\gamma}(\mathbf{\Sigma}_{\mat}) = \diag(\mathcal{S}_{\gamma}(\sval{\mat}{i}))$, and $\sval{\mat}{i}$ is the $i$-th singular value of $\mat$.
Therefore, the proximity operator of $\gamma_2(\NN{\cdot})^{*}$ in Step~\ref{al_z} of Algorithm~\ref{algo:GDPG} can be calculated with Eq.~\eqref{eqDCh} for $\gamma = \gamma_2$ and Eq.~\eqref{eq:prox_NN1} with $\gamma$ replaced by $1/\gamma_2$.

\begin{rem}[Convergence of the sequence generated by our algorithm]
    \rm{Note that $\prox_{\gamma_1(\fsamp + \iota_{\sampMSet})}(\mat)$ always returns a variable whose Frobenius norm is at most $\paramLBall$ or whose elements are between $\lbound$ and $\ubound$.
    This indicates that the sequence $\{\sampM^{(\Step)}\}_{\Step \in \mathbb{N}}$ generated by Algorithm~\ref{algo:GDPG} is bounded.
    Similarly, $\{\DCw^{(\Step)}\}_{\Step \in \mathbb{N}}$ is bounded since $\DCw^{(\Step+1)}$ lies in the unit ball of the spectral norm, whose indicator function is $(\NN{\cdot})^{*}$.
    The objective function of Prob.~\eqref{prob:proposed} is bounded below since $\fsampM \geq 0$ and $\NN{\varM \sampM}$ is bounded on the compact set $\sampMSet$.
    Since $f_1 = 0$, its gradient is $1/\beta$-Lipschitz continuous for any $\beta > 0$, and thus any $\gamma_1, \gamma_2 > 0$ satisfy the conditions in Theorem~\ref{theo:convergence_GDPGDC}.
    Finally, $\Phi$ for Prob.~\eqref{prob:proposed} is a K{\L} function~\cite{bolte2014proximal}, since $\fsampM$ is a polynomial or the $\ell_1$ norm, $\sampMSet$ and the unit ball of the spectral norm are described by polynomial inequalities, and $\langle \DCw, \varM \sampM \rangle$ is a polynomial.
    Therefore, from Theorem~\ref{theo:convergence_GDPGDC}, $\{\sampM^{(\Step)}\}_{\Step \in \mathbb{N}}$ is guaranteed to converge to a critical point of Prob.~\eqref{prob:proposed}.}
\end{rem}

\begin{rem}[Computational complexity]
    \rm{
        The computational cost of Algorithm~\ref{algo:GDPG} is determined by that of each iteration.
        The computational cost per iteration is $O(NMJ)$, where $J$ is the number of rows of $\varM$, i.e., $\varM \in \mathbb{R}^{J \times \numv}$.
        In the unconstrained case, $J = K$ for the subspace prior and $J = N$ for the smoothness and stochastic priors; thus, the computational cost per iteration is $O(NMK)$ and $O(N^2 M)$, respectively.
    }
\end{rem}

\begin{rem}[Error analysis]
    \normalfont
    \looseness=-1 Strictly speaking, the full rankness of $\varM \sampM$ is not guaranteed since we relax the constraint by minimizing $-\NN{\varM \sampM}$ as in Prob.~\eqref{prob:org4}.
    However, by minimizing $-\NN{\varM \sampM}$, our proposed DC optimization effectively pushes the singular values of $\varM \sampM$ away from zero, which inherently promotes the full rank property required for the theoretical bounds.
    In fact, $\varM \sampM$ has full rank in all cases in the experiments, so we assume that $\varM \sampM$ has full rank in the following discussion about the error bounds.
    Supposing that the full rank condition holds, the error bounds of each prior in the unconstrained case without noise can be summarized as follows:
    \begin{itemize}[align=parleft, left=1.1mm, labelsep=1.1mm, itemindent=0mm, itemsep=4.8pt, topsep=0pt, partopsep=0pt]
    \item Subspace prior: The recovery error can always be zero as:
        $\|\mathbf{x} - \tilde{\mathbf{x}}\|_2 = 0$.
    \item Smoothness prior: The recovery error is upper bounded by the signal variation bound $\smbound$ and the minimum singular value of the invertible operator $\smopeM$:
        $\|\mathbf{x} - \tilde{\mathbf{x}}\|_2 \leq \smbound / \sigma_{\min}(\smopeM)$.
    \item Stochastic prior: The recovery performance is quantified by the MSE, which depends on the covariance matrix $\stsigM$:
        $\mathbb{E}[\|\mathbf{x} - \tilde{\mathbf{x}}\|_2^2] \leq \mathrm{tr}(\stsigM)$.
\end{itemize}
\end{rem}

\section{Experiments and Results}
\label{sec:experiments}
    \looseness=1 We demonstrate the effectiveness of our method through sampling and recovery experiments across various types of synthetic graph signals and real-world data in the unconstrained case.
    All experiments were conducted using MATLAB (R2024b) on a Windows 11 computer with an Intel Core i9-14900KF 3.20-GHz processor, 32 GB of RAM, and an NVIDIA GeForce RTX 4090 GPU.
    Our method was compared with the following graph signal sampling methods: NLPD~\cite{chen2015discrete}, SASB~\cite{hara2021design}, AVM~\cite{jayawant2022practical}, GSAO~\cite{wang2023fast}, GSSS~\cite{hara2022gsss}, SUST~\cite{hara2023graph}, and a randomly designed sampling operator (RAND).
    NLPD, AVM, and GSAO are representative methods for sampling bandlimited graph signals.
    SASB is a method for sampling graph signals under only the subspace prior, which designs an aggregation sampling operator.
    GSSS is a method for sampling graph signals under arbitrary priors in the graph vertex domain with a vertex-wise strategy.
    SUST is a method for sampling graph signals under arbitrary priors in the graph frequency domain.
    As for RAND, we generated a sampling operator as a random matrix with i.i.d. entries, scaled to satisfy the same Frobenius norm bound as our Design (i), and used the same sampling and recovery process as in Eq.~\eqref{gsig_recovery}.

\subsection{Synthetic Graph Signals}
\vspace{2bp}
\subsubsection{Setup}
\begin{publicationdisplaypadding}{3pt}
    We generated random sensor graphs, which are implemented by $k$-nearest neighbor weighted graphs, whose vertices are randomly distributed in 2-D space $[0, 1] \times [0, 1]$, consisting of $\numv = 256$ vertices by using GSPBox~\cite{perraudin2014gspbox}, and also generated Erd\H{o}s--R\'enyi graphs with randomly distributed $\numv = 256$ vertices in 2-D space $[0, 1] \times [0, 1]$ and edge probability $p = 0.03$ with weights on the edges calculated by the Euclidean distance between connected vertices.
    The size of the sampled signal was set to $\nums = 16$.

    We generated the following types of graph signals.
    \begin{itemize}[align=parleft, left=1.1mm, labelsep=1.1mm, itemindent=0mm, topsep=0pt, partopsep=0pt]
        \item Subspace prior (SB)
        \begin{itemize}[align=parleft, left=0.5mm, labelsep=1.1mm, itemindent=0mm, topsep=0pt, partopsep=0pt]
            \item Bandlimited (BL) graph signals~\cite{tanaka2020sampling} characterized as:
            \begin{equation} \label{BL}
                \gsig = \sum_{i=1}^{\sbgenn} d_i \geigvec_i = \sbuniBL \sbexpd,
            \end{equation}
            where $\sbuniBL = [\geigvec_1 \ \dots \ \geigvec_{\sbgenn}]$ is the submatrix of $\guniM$ whose columns are indexed by $\BLset = \{1,\dots,\sbgenn \}$.
            Here, the generator matrix is $\sbgenM = \sbuniBL$;
            \item Periodic graph spectrum (PGS) graph signals~\cite{tanaka2020gensamp} that assume the periodicity of the graph spectrum as follows:
            \begin{equation} \label{PGS}
                \gsig = \guniM \sbgsr \DsampT \sbexpd,
            \end{equation}
            where $\sbgsr$ is a graph spectral response of the generator, which is a diagonal matrix with the $i$-th element $A(\lambda_{i}) = \exp(-1.5 \lambda_{i}/\lambda_{\mathrm{max}})$, where $\lambda_{i}$ and $\lambda_{\mathrm{max}}$ are the $i$-th graph frequency and the largest graph frequency, respectively, $\Dsamp = [ \onesK \ \onesK \ \dots ] \in \realKN$ is the matrix for the GFT domain upsampling, and $\onesK$ is the $\sbgenn \times \sbgenn$ identity matrix.
            In this case, the generator matrix is $\sbgenM = \guniM \sbgsr \DsampT$;
            \item Piecewise constant (PWC) graph signals~\cite{chen2016representations} that are characterized by constant values in separated vertex regions and are defined as follows with the number of pieces $\sbgenn$:
            \begin{equation} \label{PWC}
                \gsig = \sum_{i=1}^{\sbgenn} d_i \sbpwc{i} = [\sbpwc{1} \ \dots \ \sbpwc{\sbgenn}] \sbexpd,
            \end{equation}
            where $\sbpwc{i} \in \realN$ for any $i = 1,\dots,\sbgenn$ is defined as $\sbelepwc{i}{j} = 1$ when vertex $j$ is in the $i$-th piece and $\sbelepwc{i}{j} = 0$ otherwise for any $j = 1,\dots,\numv$.
            In this case, the generator matrix is $\sbgenM = [\sbpwc{1} \ \dots \ \sbpwc{\sbgenn}]$.
        \end{itemize}
        For the signal types under the subspace prior, we drew each element $d_i$ of $\sbexpd$ from $\mathcal{N}(1,1)$ for all $i = 1,\dots,\sbgenn$, and set $\sbgenn = \nums$.
        \item Smoothness prior (SM)
            \begin{itemize}[align=parleft, left=0.5mm, labelsep=1.1mm, itemindent=0mm, topsep=0pt, partopsep=0pt]
            \item Gaussian Markov random field (GMRF)~\cite{gadde2015probabilistic} graph signals with the power spectrum for any $i = 1, \ldots, \numv$:
            \begin{equation}
                \smps = 0.1 / \left(\lambda_{i} + 0.1\right);
            \end{equation}
            \item Piecewise linear (PWL) signals with $8$ randomly chosen vertices having values based on uniformly distributed random numbers in $[-1,1]$ and interpolated values on other vertices based on the topology of the graph for each intervening vertex.
        \end{itemize}
        \item Stochastic prior (ST)
            \begin{itemize}[align=parleft, left=0.5mm, labelsep=1.1mm, itemindent=0mm, topsep=0pt, partopsep=0pt]
            \item Stochastic graph signals (SGS), i.e., Gaussian graph signals with the band-pass power spectrum for any $i = 1, \ldots, \numv$:
            \begin{equation}
                \stelefsigM = \exp \left( - \left(
                    \frac{
                        {2 \lambda_{i} - \lambda_{\max}}
                    }{
                        {\sqrt{\lambda_{\max}}}
                    }
                \right) ^2\right).
            \end{equation}
        \end{itemize}
    \end{itemize}

    For all signal types above, we also experimented with noisy sampled signals $\ssign := \gsampsig + \stnoise$, where $\stnoise \in \realM$ is generated as zero-mean i.i.d. Gaussian noise with variance $\sigma^2=0.3$.

    \PublicationWide{421.7bp}{%
  \begin{table*}[!t]
    \captionsetup{font=footnotesize}
    \centering
    \caption{ \footnotesize
        Average MSEs in Decibels for 20 Independent Runs on Random Sensor Graphs.
    }
    \vspace{0mm}
    \label{table:M16}
    \footnotesize
    \begin{minipage}{1\textwidth}
        \centering
        \setlength{\tabcolsep}{4pt}
        {\fontsize{7}{7.5}\selectfont
            \begin{tabular}{@{}c@{\hspace{3pt}}c@{\hspace{3pt}}cccccccccc}
            \toprule
                \multirow{2}{*}[-1mm]{Priors} & \multirow{2}{*}[-1mm]{\shortstack{Signal\\Types}} & \multicolumn{10}{c}{Methods} \\
                \cmidrule(lr){3-12}
                && NLPD~\cite{chen2015discrete} & SASB~\cite{hara2021design} & AVM~\cite{jayawant2022practical} & GSAO~\cite{wang2023fast} & GSSS~\cite{hara2022gsss} & SUST~\cite{hara2023graph} & RAND & \textbf{Ours (i)} & \textbf{Ours (ii)} & \textbf{Ours (iii)}\\
            \midrule
            \midrule
                \multirow{12}{*}[-2mm]{SB}
                    & \multirow{2}{*}[0mm]{BL}
                        & $-606.576$ & $-325.615$ & $-485.269$ & $\underline{-609.114}$ & $-609.070$ & $-590.288$ & $-570.978$ & $-608.435$ & $\mathbf{-610.582}$ & $-606.854$  \\
                        && $  (-603.953) $ & $  (-316.423) $ & $  (-472.514) $ & $  (-607.724) $ & $  (-615.953) $ & $  (-590.005) $ & $(-560.996)$ & $  (-609.112) $ & $  (-613.396) $ & $  (-609.124) $ \\
                    & \multirow{2}{*}{+noise}
                        & $-11.938$ & $10.702$ & $-3.663$ & $-11.283$ & $-11.212$ & $-34.211$ & $25.995$ & $\mathbf{-58.269}$ & $-33.115$ & $\underline{-48.895}$ \\
                        && $  (-17.621) $ & $  (10.597) $ & $  (-3.271) $ & $  (-17.313) $ & $  (-17.226) $ & $  (-41.962) $ & $(38.338)$ & $  (-66.039) $ & $  (-40.512) $ & $  (-50.203) $ \\

                \cmidrule(lr){2-12}
                    & \multirow{2}{*}[0mm]{PGS}
                        & $4.236$ & $-570.746$ & $13.900$ & $4.520$ & $\underline{-593.450}$ & $\mathbf{-595.164}$ & $-567.382$ & $-592.823$ & $-587.593$ & $-587.042$ \\
                        && $  (-2.444) $ & $  (-567.012) $ & $  (21.697) $ & $  (-2.453) $ & $  (-598.757) $ & $  (-595.047) $ & $(-562.714)$ & $  (-591.918) $ & $  (-589.290) $ & $  (-587.843) $ \\
                    & \multirow{2}{*}{+noise}
                        & $6.094$ & $22.727$ & $11.496$ & $5.757$ & $-10.964$ & $\mathbf{-63.596}$ & $2.266$ & $-57.484$ & $-42.043$ & $\underline{-62.902}$  \\
                        && $  (0.304) $ & $  (29.019) $ & $  (14.643) $ & $  (-0.930) $ & $  (-16.043) $ & $  (-72.210) $ & $(7.194)$ & $  (-66.494) $ & $  (-50.259) $ & $  (-72.160) $ \\

                \cmidrule(lr){2-12}
                    & \multirow{2}{*}[0mm]{PWC}
                        & $-4.756$ & $-589.948$ & $6.508$ & $-3.974$ & $\mathbf{-590.645}$ & $-513.194$ & $-556.180$ & $-588.781$ & $\underline{-590.061}$ & $-583.532$  \\
                        && $  (-9.263) $ & $  (-586.164) $ & $  (7.896) $ & $  (-8.160) $ & $  (-587.706) $ & $  (-500.984) $ & $(-547.731)$ & $  (-585.517) $ & $  (-589.406) $ & $  (-584.062) $ \\
                    & \multirow{2}{*}{+noise}
                        & $-2.147$ & $-10.925$ & $13.664$ & $-1.471$ & $-12.630$ & $69.739$ & $11.338$ & $-57.774$ & $\underline{-58.170}$ & $\mathbf{-66.172}$  \\
                        && $  (-10.412) $ & $  (-20.655) $ & $  (16.386) $ & $  (-10.136) $ & $  (-21.108) $ & $  (82.732) $ & $(18.157)$ & $  (-66.539) $ & $  (-67.468) $ & $  (-70.869) $  \\

            \midrule
                \multirow{8}{*}[-1mm]{SM}
                    & \multirow{2}{*}[0mm]{GMRF}
                        & $-14.806$ & $-$ & $5.968$ & $-14.779$ & $-18.233$ & $-18.907$ & $-18.522$ & $\mathbf{-21.281}$ & $\underline{-20.668}$ & $-20.338$ \\
                        && $  (-26.682) $ & $-$ & $  (12.492) $ & $  (-26.410) $ & $  (-33.307) $ & $  (-34.761) $ & $(-34.638)$ & $  (-36.492) $ & $  (-35.435) $ & $  (-36.566) $ \\
                    & \multirow{2}{*}{+noise}
                        & $-6.445$ & $-$ & $11.749$ & $-5.924$ & $-14.888$ & $-18.790$ & $-18.209$ & $\mathbf{-21.147}$ & $\underline{-20.621}$ & $-20.313$ \\
                        && $  (-16.669) $ & $-$ & $  (16.203) $ & $  (-14.891) $ & $  (-32.288) $ & $  (-34.920) $ & $(-34.638)$ & $  (-36.544) $ & $  (-35.437) $ & $  (-36.620) $ \\

                \cmidrule(lr){2-12}
                    & \multirow{2}{*}[0mm]{PWL}
                        & $\underline{-62.276}$ & $-$  & $-37.465$ & $-58.850$ & $-34.056$ & $-35.545$ & $-34.566$ & $\mathbf{-67.391}$ & $-52.062$ & $-43.786$  \\
                        && $  (-63.450) $ & $-$  & $  (-30.771) $ & $  (-57.289) $ & $  (-37.036) $ & $  (-39.100) $ & $(-39.047)$ & $  (-67.945) $ & $  (-51.864) $ & $  (-45.161) $ \\
                    & \multirow{2}{*}{+noise}
                        & $-11.772$ & $-$ & $14.800$ & $-11.956$ & $-22.281$ & $-34.814$ & $-32.613$ & $\mathbf{-55.387}$ & $\underline{-50.491}$ & $-43.323$ \\
                        && $  (-19.009) $ & $-$ & $  (20.634) $ & $  (-20.401) $ & $  (-31.720) $ & $  (-39.065) $ & $(-38.113)$ & $  (-64.019) $ & $  (-51.866) $ & $  (-45.193) $ \\

            \midrule
                \multirow{4}{*}[0mm]{ST}
                    & \multirow{2}{*}[0mm]{SGS}
                        & $-2.998$ & $-$ & $6.618$ & $-3.434$ & $-8.582$ & $-8.933$ & $-8.995$ & $\mathbf{-9.378}$ & $\underline{-9.348}$ & $-9.129$ \\
                        && $  (-14.488) $ & $-$ & $  (7.422) $ & $  (-15.164) $ & $  (-25.563) $ & $  (-27.131) $ & $(-25.590)$ & $  (-26.235) $ & $  (-26.756) $ & $  (-25.624) $ \\
                    & \multirow{2}{*}{+noise}
                        & $-0.386$ & $-$ & $15.956$ & $-0.463$ & $-8.017$ & $-8.837$ & $-8.940$ & $\mathbf{-9.352}$ & $-8.860$ & $\underline{-9.103}$ \\
                        && $  (-10.655) $ & $-$ & $  (21.047) $ & $  (-12.100) $ & $  (-25.113) $ & $  (-27.096) $ & $(-25.500)$ & $  (-26.225) $ & $  (-26.702) $ & $  (-25.642) $ \\
            \bottomrule
            \end{tabular}
        }
        \\ \vspace*{1mm}
        \small
        \raggedright
    \end{minipage}
    \vspace*{-3mm}
\end{table*}
\vspace{29.85bp}\def\resultA{$\underline{-52.324}$}
\def\resultB{$-17.603$}
\def\resultC{$-52.181$}
\def\resultD{$-35.036$}
\def\resultE{$-33.830$}
\def\resultF{$-30.922$}
\def\resultG{$\mathbf{-55.155}$}
\def\resultH{$-49.284$}
\def\resultI{$-43.881$}

\def\resultNoiseA{$-10.219$}
\def\resultNoiseB{$30.637$}
\def\resultNoiseC{$-10.543$}
\def\resultNoiseD{$-20.045$}
\def\resultNoiseE{$-32.793$}
\def\resultNoiseF{$-29.534$}
\def\resultNoiseG{$\mathbf{-49.554}$}
\def\resultNoiseH{$\underline{-47.876}$}
\def\resultNoiseI{$-43.286$}

\newcounter{subsubfigure}

\newcommand{\subsubcaption}[2]{%
    \backsubcaption{#1}
    \stepcounter{subsubfigure}%
    \subcaption[\alph{subfigure}-\arabic{subsubfigure}]{#2}%
    \resetsubsubcaption
}

\newcommand{\backsubcaption}[1]{%
    \addtocounter{subfigure}{-#1}%
}

\newcommand{\resetsubsubcaption}{%
    \setcounter{subsubfigure}{1}%
}

\begin{figure*}[t]
    \centering
    \captionsetup{font=footnotesize}
    \captionsetup[subfigure]{font=scriptsize}

    \begin{subfigure}[t]{.091\textwidth}
        \includegraphics[width=\linewidth]{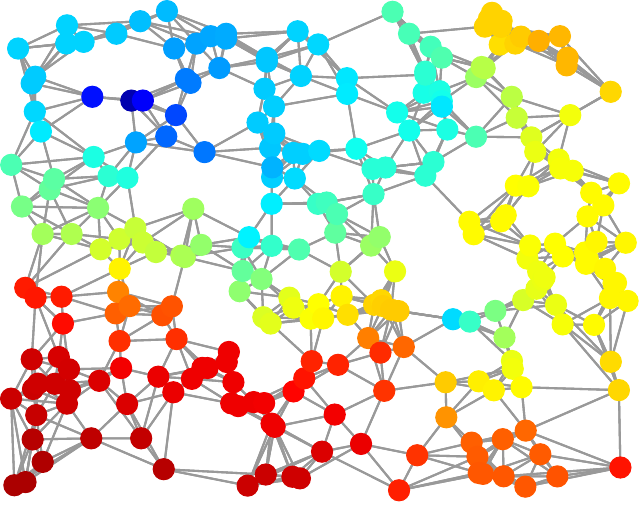}
        \vspace*{-6mm}
        \subcaption{Original \\ MSE [dB] }
    \end{subfigure}
    \renewcommand{\thesubfigure}{\alph{subfigure}-\arabic{subsubfigure}}
    \hfill
    \begin{subfigure}[t]{.091\textwidth}
        \includegraphics[width=\linewidth]{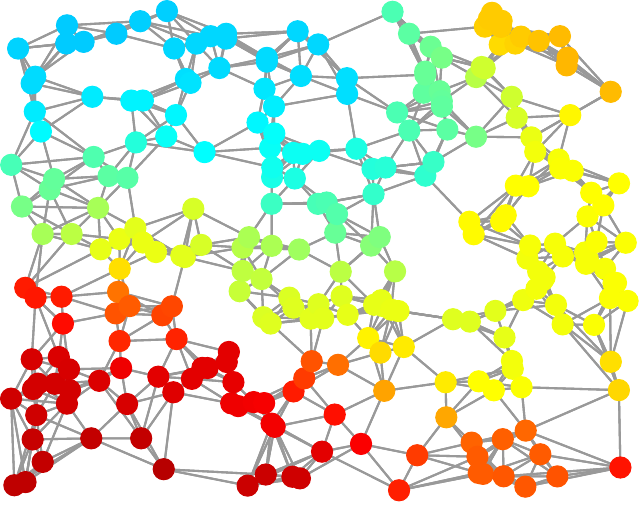}
        \vspace*{-6mm}
        \resetsubsubcaption
        \subcaption{NLPD \\ \resultA }
    \end{subfigure}
    \hfill
    \begin{subfigure}[t]{.091\textwidth}
        \includegraphics[width=\linewidth]{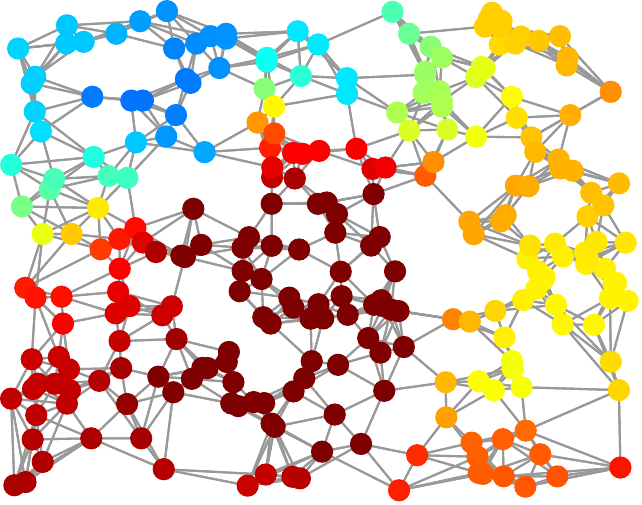}
        \vspace*{-6mm}
        \subcaption{AVM \\ \resultB }
    \end{subfigure}
    \hfill
    \begin{subfigure}[t]{.091\textwidth}
        \includegraphics[width=\linewidth]{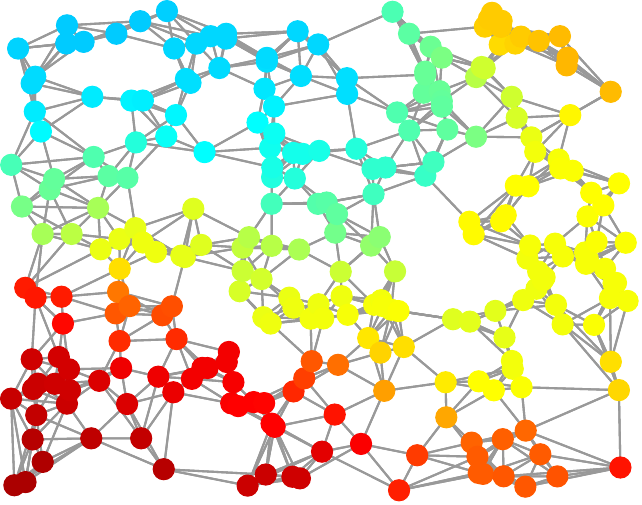}
        \vspace*{-6mm}
        \subcaption{GSAO \\ \resultC }
    \end{subfigure}
    \hfill
    \begin{subfigure}[t]{.091\textwidth}
        \includegraphics[width=\linewidth]{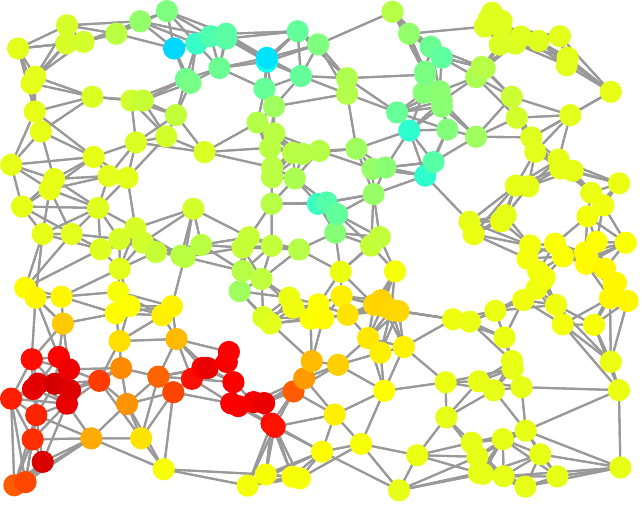}
        \vspace*{-6mm}
        \subcaption{GSSS \\ \resultD }
    \end{subfigure}
    \hfill
    \begin{subfigure}[t]{.091\textwidth}
        \includegraphics[width=\linewidth]{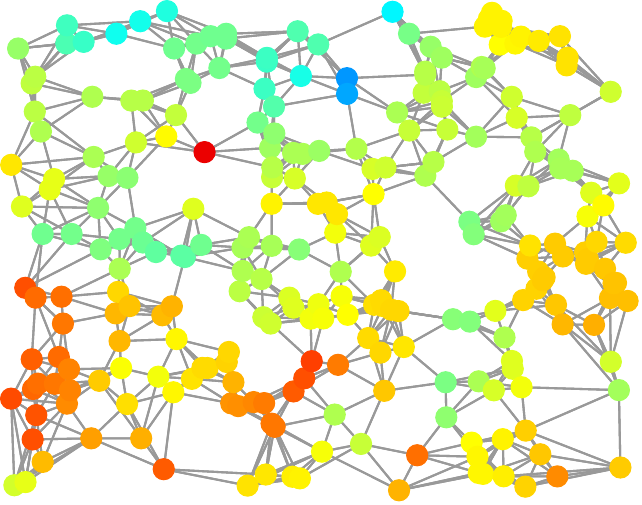}
        \vspace*{-6mm}
        \subcaption{SUST \\ \resultE}
    \end{subfigure}
    \hfill
    \begin{subfigure}[t]{.091\textwidth}
        \includegraphics[width=\linewidth]{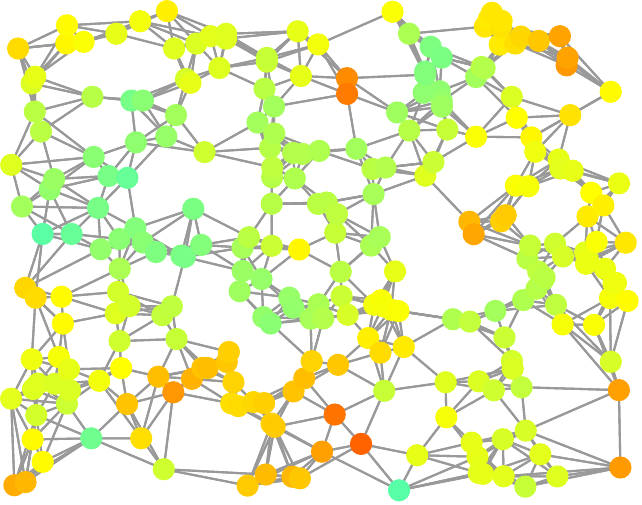}
        \vspace*{-6mm}
        \subcaption{RAND \\ \resultF}
    \end{subfigure}
    \hfill
    \begin{subfigure}[t]{.091\textwidth}
        \includegraphics[width=\linewidth]{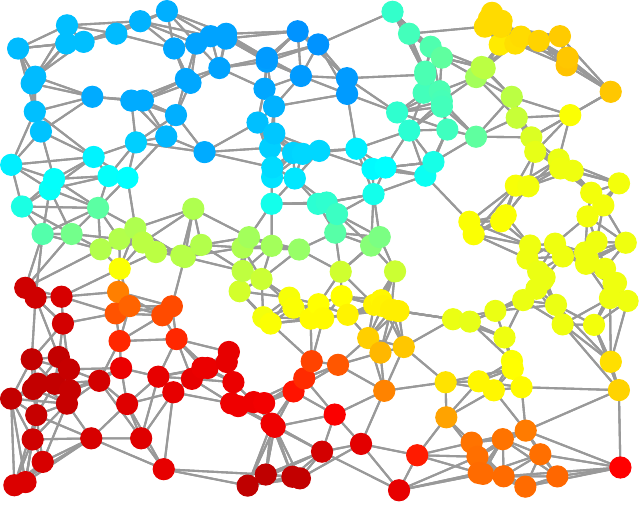}
        \vspace*{-6mm}
        \subcaption{\textbf{Ours (i)} \\ \resultG}
    \end{subfigure}
    \hfill
    \begin{subfigure}[t]{.091\textwidth}
        \includegraphics[width=\linewidth]{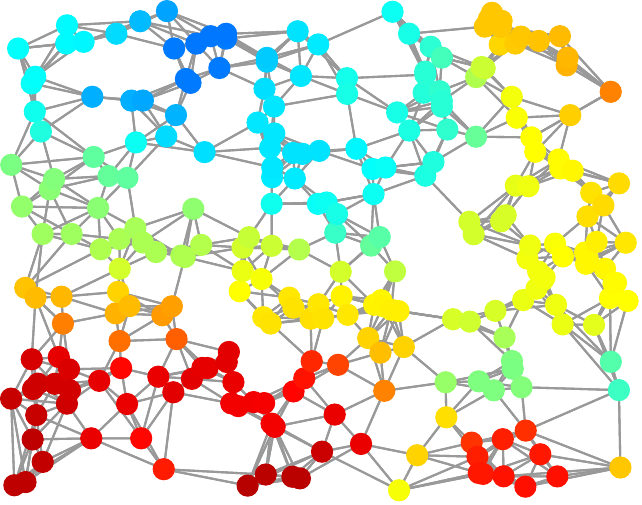}
        \vspace*{-6mm}
        \subcaption{\textbf{Ours (ii)} \\ \resultH}
    \end{subfigure}
    \hfill
    \begin{subfigure}[t]{.091\textwidth}
        \includegraphics[width=\linewidth]{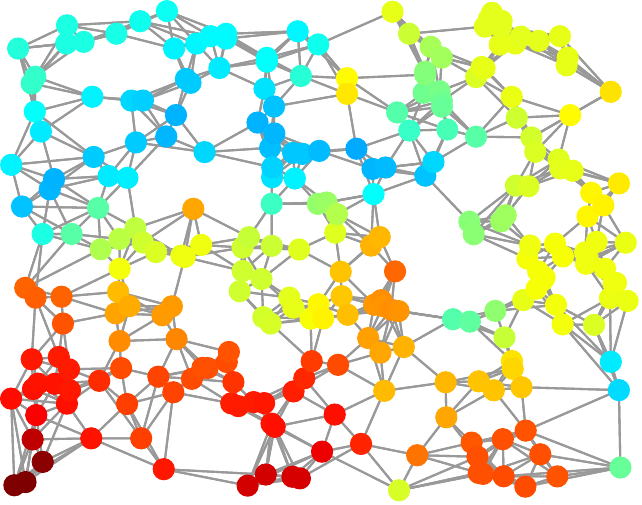}
        \vspace*{-6mm}
        \subcaption{\textbf{Ours (iii)} \\ \resultI}
    \end{subfigure}
    \hfill
        \begin{subfigure}{.017\textwidth}
        \includegraphics[width=\linewidth]{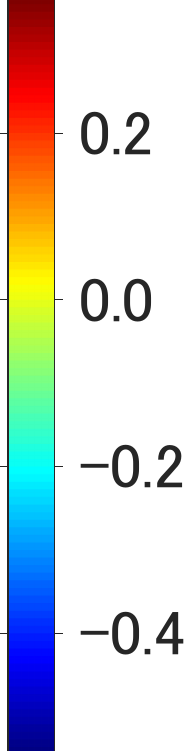}
    \end{subfigure}

    \begin{subfigure}[t]{.091\textwidth}
        \hspace*{\fill}
    \end{subfigure}
    \hfill
    \begin{subfigure}[t]{.091\textwidth}
        \includegraphics[width=\linewidth]{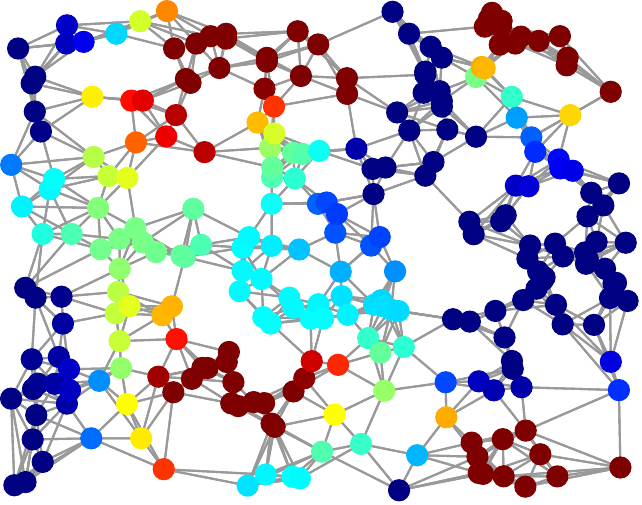}
        \vspace*{-6mm}
        \subsubcaption{9}{NLPD \\ \resultNoiseA }
    \end{subfigure}
    \hfill
    \begin{subfigure}[t]{.091\textwidth}
        \includegraphics[width=\linewidth]{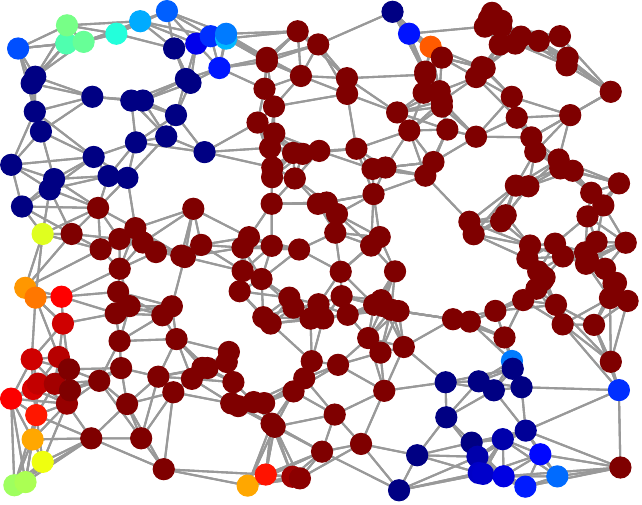}
        \vspace*{-6mm}
        \subsubcaption{0}{AVM \\ \resultNoiseB }
    \end{subfigure}
    \hfill
    \begin{subfigure}[t]{.091\textwidth}
        \includegraphics[width=\linewidth]{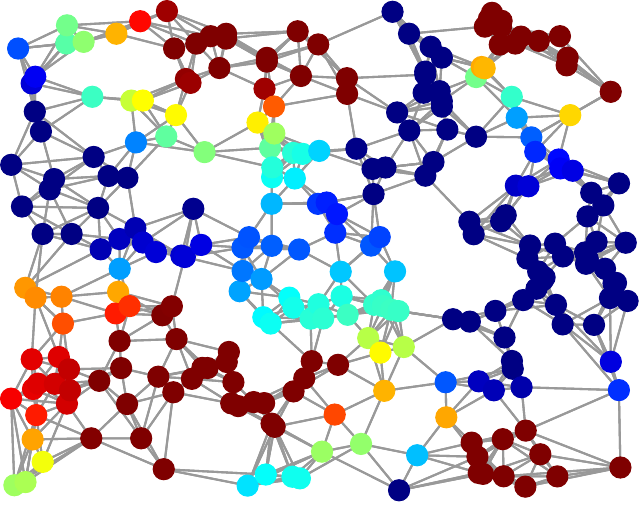}
        \vspace*{-6mm}
        \subsubcaption{0}{GSAO \\ \resultNoiseC }
    \end{subfigure}
    \hfill
    \begin{subfigure}[t]{.091\textwidth}
        \includegraphics[width=\linewidth]{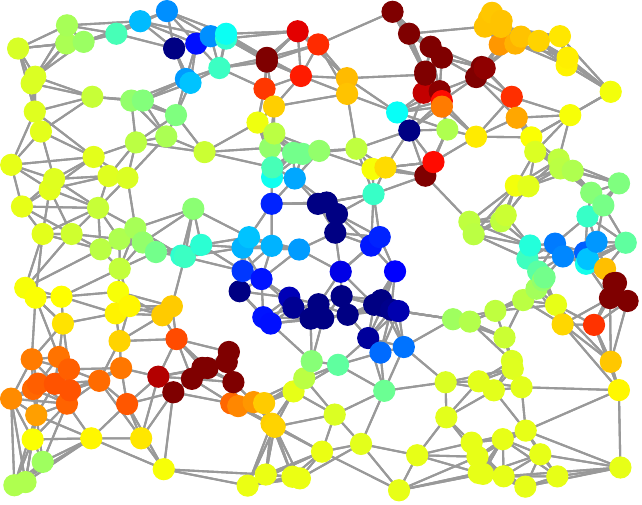}
        \vspace*{-6mm}
        \subsubcaption{0}{GSSS \\ \resultNoiseD }
    \end{subfigure}
    \hfill
    \begin{subfigure}[t]{.091\textwidth}
        \includegraphics[width=\linewidth]{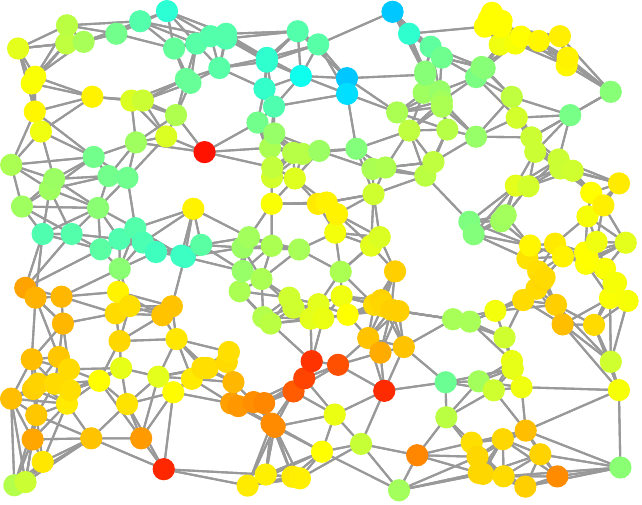}
        \vspace*{-6mm}
        \subsubcaption{0}{SUST \\ \resultNoiseE}
    \end{subfigure}
    \hfill
    \begin{subfigure}[t]{.091\textwidth}
        \includegraphics[width=\linewidth]{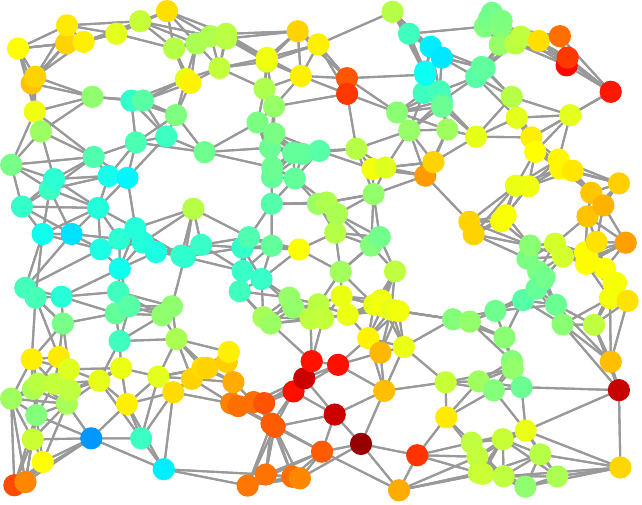}
        \vspace*{-6mm}
        \subsubcaption{0}{RAND \\ \resultNoiseF}
    \end{subfigure}
    \hfill
    \begin{subfigure}[t]{.091\textwidth}
        \includegraphics[width=\linewidth]{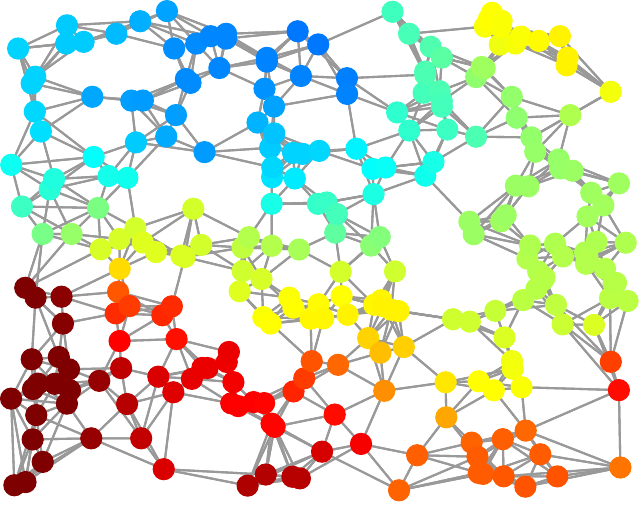}
        \vspace*{-6mm}
        \subsubcaption{0}{\textbf{Ours (i)} \\ \resultNoiseG}
    \end{subfigure}
    \hfill
    \begin{subfigure}[t]{.091\textwidth}
        \includegraphics[width=\linewidth]{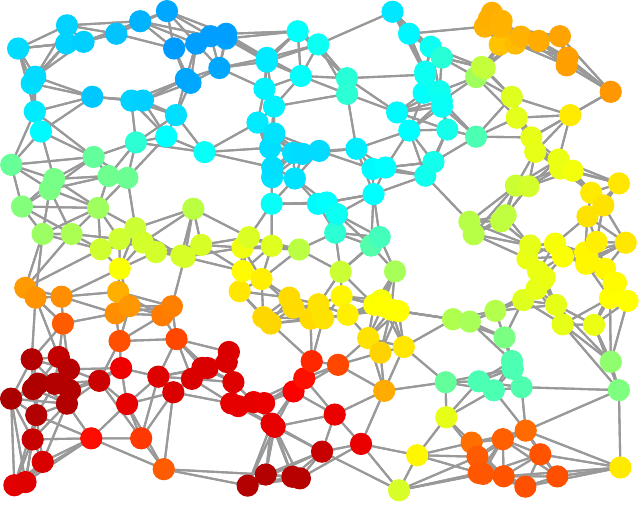}
        \vspace*{-6mm}
        \subsubcaption{0}{\textbf{Ours (ii)} \\ \resultNoiseH}
    \end{subfigure}
    \hfill
    \begin{subfigure}[t]{.091\textwidth}
        \includegraphics[width=\linewidth]{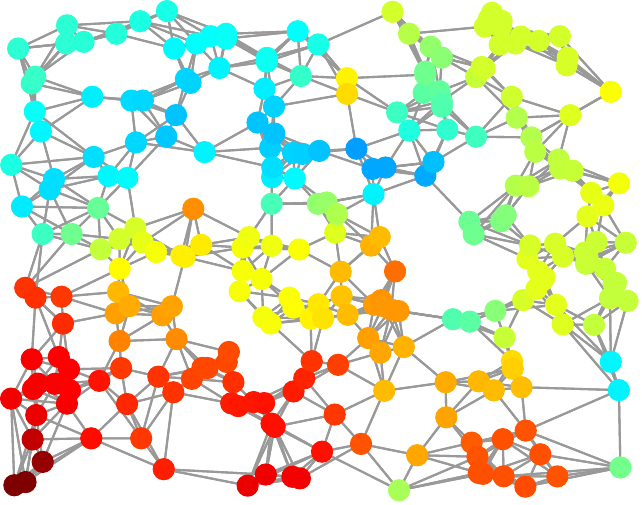}
        \vspace*{-6mm}
        \subsubcaption{0}{\textbf{Ours (iii)} \\ \resultNoiseI}
    \end{subfigure}
    \hfill
    \begin{subfigure}{.017\textwidth}
        \includegraphics[width=\linewidth]{12_fig/fig/PWL/01withoutNoise/colorbar.pdf}
    \end{subfigure}

    \vspace*{-1mm}
    \caption{
        An example of a PWL graph signal defined on a sensor graph and its sampled and recovered signals using each method.
        (a) shows the original PWL graph signal.
        (b-1)--(j-1) show the results without noise.
        (b-2)--(j-2) show the results with noise added to the sampled signals.
    }
\label{fig:results_PWL_UN}
\vspace{0mm}
\end{figure*}
}
    \looseness=1 As for the parameters, $\paramLBall$ in~\eqref{prob:Lball} was set as $\paramLBall = \sqrt{\numv\nums}/4$.
    The parameters $\lbound$ and $\ubound$ in~\eqref{prob:FNandBox} and~\eqref{prob:L1andBox} were set as $\lbound = 0$ and $\ubound = 1$.
    The parameter $\paramBox$ in~\eqref{prob:FNandBox} was set as $\paramBox = 0.5$, and $\paramBox$ in~\eqref{prob:L1andBox} was set as $\paramBox = 0.1$.
    For our proposed algorithm, $\gamma_1$ and $\gamma_2$ were set as $\gamma_1 = \gamma_2 = 0.001$ for all designs.
    We defined $\sampM^{(0)}$ as a matrix with random Gaussian entries, and the stopping criterion in Algorithm~\ref{algo:GDPG} as $\|\sampM^{(\Step + 1)} - \sampM^{(\Step)}\|_{F}/\|\sampM^{(\Step)}\|_{F} \leq 10^{-5}$.
    The parameter settings for the existing methods followed the descriptions in the respective papers.
    For sampling graph signals under the smoothness prior, we used the spectral response of $\smopeM$ as $F(\lambda_i) = (\lambda_i / \lambda_\mathrm{max}) + 0.1$.
    For the quantitative evaluations, we used MSE: $\mathrm{MSE} = \|\tilde{\gsig} - \gsig\|^{2}_2/\numv$.

\end{publicationdisplaypadding}
\subsubsection{Results and Discussion}

    Since the evaluation on the random sensor graphs and Erdős–Rényi graphs yielded similar trends and performance characteristics, we primarily present the results and discussion based on the random sensor graphs in this section.

    Table~\ref{table:M16} shows the averaged MSEs in decibels between original graph signals and recovered graph signals generated by each method obtained from 20 independent runs for each graph
\PublicationColumn{164bp}{\begin{table}[!t]
    \captionsetup{font=footnotesize}
    \caption{
        Average Percentage of Zero Elements in Sampling Operators Designed for Each Signal Type.
    }
    \vspace{0mm}
    \label{table:zeros}
    \footnotesize
        \centering
        \setlength{\tabcolsep}{4pt}
        {\fontsize{8}{8.2}\selectfont
            \begin{tabular}{cc>{\centering\arraybackslash}p{1.4cm}>{\centering\arraybackslash}p{1.4cm}>{\centering\arraybackslash}p{1.4cm}}
            \toprule
                \multirow{2}{*}[-1mm]{Priors} & \multirow{2}{*}[-1mm]{\shortstack{Signal\\Types}} & \multicolumn{3}{c}{Methods} \\
                \cmidrule(lr){3-5}
                && Ours (i) & Ours (ii) & Ours (iii)\\
            \midrule
            \midrule
                \multirow{3}{*}[-1.5mm]{SB}
                    & \multirow{1}{*}{BL}
                        & $0.001$ & $53.588$ & $91.077$ \\
                \cmidrule(lr){2-5}
                    & \multirow{1}{*}{PGS}
                        & $0.006$ & $51.599$ & $68.799$ \\
                \cmidrule(lr){2-5}
                    & \multirow{1}{*}{PWC}
                        & $0.000$ & $61.639$ & $61.453$ \\

            \midrule
                \multirow{2}{*}[-1mm]{SM}
                    & \multirow{1}{*}{GMRF}
                        & $0.001$ & $66.783$ & $51.608$ \\
                \cmidrule(lr){2-5}
                    & \multirow{1}{*}{PWL}
                        & $0.005$ & $66.807$ & $51.267$ \\

            \midrule
                \multirow{1}{*}[0mm]{ST}
                    & \multirow{1}{*}{SGS}
                        & $0.071$ & $62.759$ & $90.770$ \\

            \bottomrule
            \end{tabular}
        }
        \\ \vspace*{1mm}
        \small
        \raggedright
    \vspace*{-1mm}
\end{table}
}
signal type.
    The numbers in parentheses in the table represent the standard deviations of the results in decibels.
    Fig.~\ref{fig:results_PWL_UN} visualizes examples of graph signals and their sampled and recovered signals by each method.
    Ours (i)--(iii) refer to Designs (i)--(iii) of the proposed method described in Section~\ref{sec:propose}, respectively.
    SASB is designed to sample only graph signals under
  \PublicationWide{122bp}{\begin{table*}[!t]
    \captionsetup{font=footnotesize}
    \centering
    \caption{
        Average Runtime in Seconds for Each Method on Random Sensor Graphs.
    }
    \vspace{0mm}
    \label{table:runtime}
    \footnotesize
    \begin{minipage}{1\textwidth}
        \centering
        \setlength{\tabcolsep}{4.5pt}
        {\fontsize{8}{8}\selectfont
            \begin{tabular}{>{\centering\arraybackslash}p{0.75cm}>{\centering\arraybackslash}p{0.95cm}ccccccccc}
            \toprule
                \multirow{2}{*}[-1mm]{Priors} & \multirow{2}{*}[-1mm]{\shortstack{Signal\\Types}} & \multicolumn{9}{c}{Methods} \\
                \cmidrule(lr){3-11}
                && NLPD~\cite{chen2015discrete} & SASB~\cite{hara2021design} & AVM~\cite{jayawant2022practical} & GSAO~\cite{wang2023fast} & GSSS~\cite{hara2022gsss} & SUST~\cite{hara2023graph} & \textbf{Ours (i)} & \textbf{Ours (ii)} & \textbf{Ours (iii)}\\
            \midrule
            \midrule
                \multirow{3}{*}{SB}
                    & BL
                        & $0.070$ & $9.925$ & $0.039$ & $0.010$ & $0.098$ & $0.037$ & $1.317$ & $5.363$ & $5.924$ \\
                    & PGS
                        & $0.065$ & $1.037$ & $0.041$ & $0.012$ & $0.100$ & $0.037$ & $0.634$ & $5.101$ & $4.153$ \\
                    & PWC
                        & $0.067$ & $1.150$ & $0.041$ & $0.012$ & $0.100$ & $0.038$ & $0.458$ & $2.688$ & $2.439$ \\
            \midrule
                \multirow{2}{*}{SM}
                    & GMRF
                        & $0.064$ & $-$ & $0.040$ & $0.011$ & $0.099$ & $0.041$ & $3.500$ & $10.628$ & $6.594$ \\
                    & PWL
                        & $0.066$ & $-$ & $0.039$ & $0.011$ & $0.099$ & $0.041$ & $4.092$ & $10.860$ & $6.499$ \\
            \midrule
                \multirow{1}{*}{ST}
                    & SGS
                        & $0.066$ & $-$ & $0.041$ & $0.011$ & $0.099$ & $0.037$ & $8.396$ & $16.464$ & $9.009$ \\
            \bottomrule
            \end{tabular}
        }
        \\ \vspace*{1mm}
        \small
        \raggedright
    \end{minipage}
    \vspace{0mm}
\end{table*}
}%
  \noindent\begin{minipage}[t][164bp][t]{\columnwidth}%
    \vspace{16bp}\PublicationContent{\begin{figure}[t]
    \centering
    \captionsetup{font=footnotesize}
    \begin{subfigure}[t]{0.05\linewidth}
        \hfill
    \end{subfigure}
    \hfill
    \begin{subfigure}[t]{0.23\linewidth}
        \includegraphics[width=\linewidth]{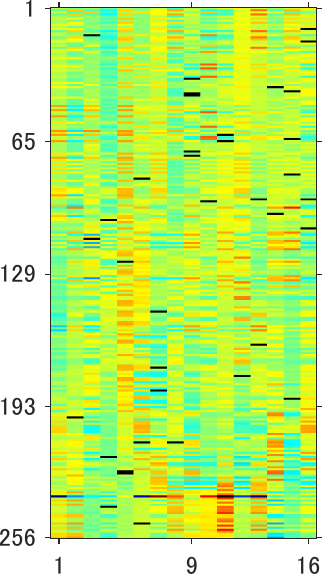}
        \vspace*{-6mm}
        \subcaption{Design (i)}
    \end{subfigure}
    \hfill
    \begin{subfigure}[t]{0.23\linewidth}
        \includegraphics[width=\linewidth]{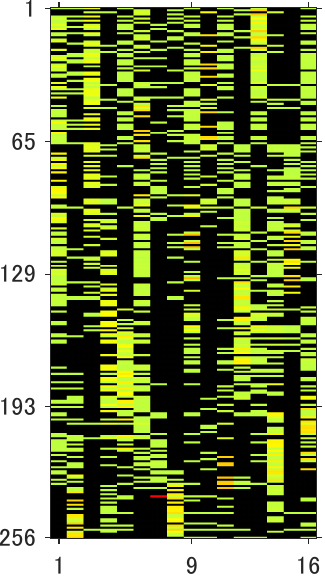}
        \vspace*{-6mm}
        \subcaption{Design (ii)}
    \end{subfigure}
    \hfill
    \begin{subfigure}[t]{0.23\linewidth}
        \includegraphics[width=\linewidth]{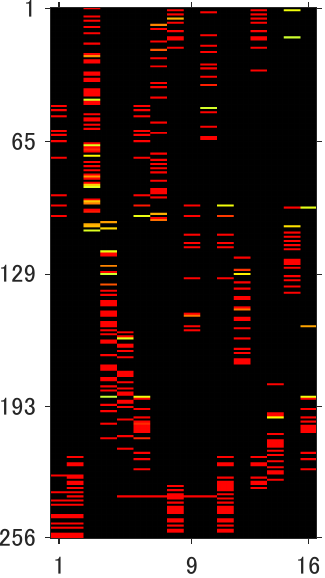}
        \vspace*{-6mm}
        \subcaption{Design (iii)}
    \end{subfigure}
    \hfill
    \begin{subfigure}[t]{0.053\linewidth}
        \raisebox{1.6mm}{\includegraphics[width=\linewidth]{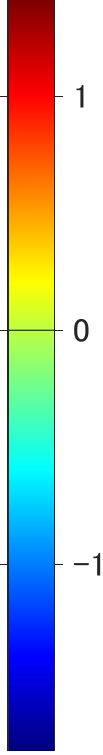}}
    \end{subfigure}
    \hfill
    \begin{subfigure}[t]{0.05\linewidth}
        \hfill
    \end{subfigure}
    \vspace*{-1mm}
    \caption{
        An example of visualized sampling operators $\sampM$ designed by the proposed method under the subspace prior.
        Zero values are shown in black.
        }
    \label{fig:results_sampM}
    \vspace{0mm}
\end{figure}
}%
  \end{minipage}\par\noindent

the subspace prior, so the results of SASB are only shown in the subspace prior rows in the table.
    The best and second best results in each case are highlighted in bold and with underline, respectively.

    From the table, our method achieves the best or second best results in most cases.
    Under the subspace prior, some of the results of existing methods are better than those of our method, but the MSEs in those cases are about $-590$ dB, which is on the order of $10^{-29}$ to $10^{-30}$, indicating that they achieve very good results in those cases.
    Comparing the standard deviations of the results in Table~\ref{table:M16}, we see that our method tends to have smaller standard deviations than existing methods in many cases, indicating that our method demonstrates stable performance.
    Those results indicate the effectiveness of our approach in designing sampling operators.

    We also compare the sampling operators designed by the proposed method.
    Table~\ref{table:zeros} shows the average percentage of zero elements (absolute value less than $10^{-5}$) in the designed sampling operators for each signal type.
    Fig.~\ref{fig:results_sampM} visualizes the sampling operators designed by our method for a bandlimited graph signal in the experiment.
    Fig.~\ref{fig:results_plotSampM} visualizes the filter-like aggregation weights in the first column of the sampling operators shown in Fig.~\ref{fig:results_sampM}, which corresponds to the weights for generating the first element of the sampled signal.

    Although Design (i) imposes a constraint to keep the Frobenius norm of $\sampM$ at most $\paramLBall$, it does not impose any constraints on the individual elements of $\sampM$.
    Hence, it tends to have nonzero values evenly across its elements, as shown in Table~\ref{table:zeros} and visualized in Figs.~\ref{fig:results_sampM} and~\ref{fig:results_plotSampM}, resulting in utilizing signal values from all vertices for sampling.

    As for Design (iii), this design imposes an $\ell_1$-norm penalty, which encourages sparsity in $\sampM$.
    Thus, as shown in Table~\ref{table:zeros},
\PublicationColumn{308bp}{%
  \vspace{14.88bp}\begin{figure}[t]
    \centering
    \captionsetup{font=footnotesize}
    \begin{subfigure}[t]{0.32\linewidth}
        \includegraphics[width=\linewidth]{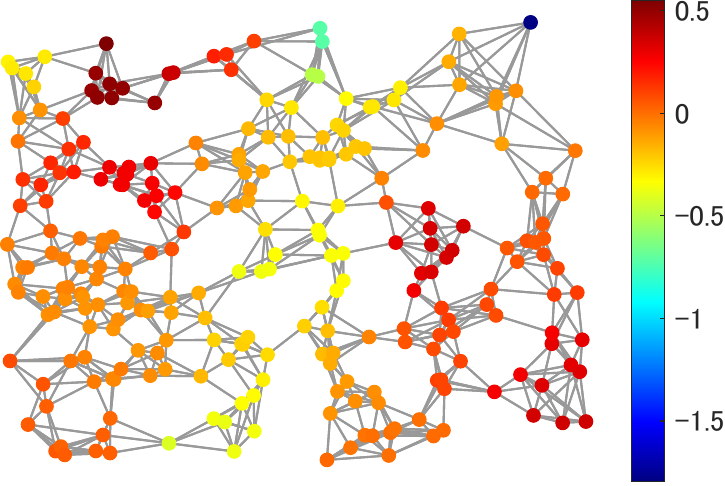}
        \vspace*{-6mm}
        \subcaption{Design (i)}
    \end{subfigure}
    \hfill
    \begin{subfigure}[t]{0.32\linewidth}
        \includegraphics[width=\linewidth]{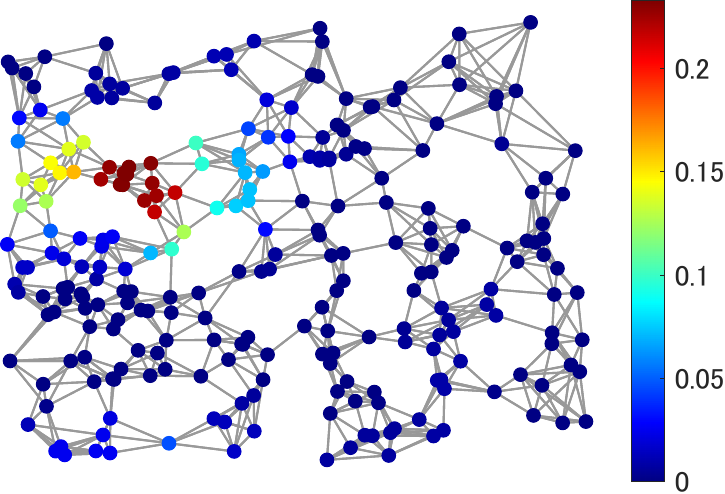}
        \vspace*{-6mm}
        \subcaption{Design (ii)}
    \end{subfigure}
    \hfill
    \begin{subfigure}[t]{0.32\linewidth}
        \includegraphics[width=\linewidth]{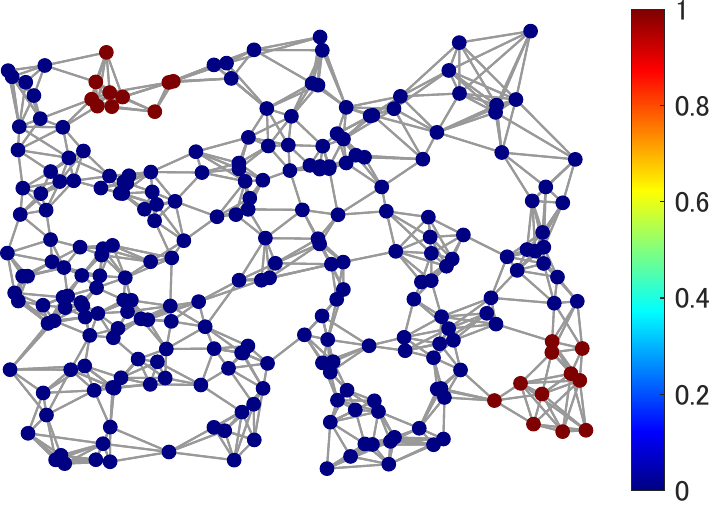}
        \vspace*{-6mm}
        \subcaption{Design (iii)}
    \end{subfigure}
    \vspace*{-1mm}
    \caption{
        An example of filter-like aggregation weights corresponding to the first row of the sampling operator $\sampMT$ shown in Fig.~\ref{fig:results_sampM}.
    }
    \label{fig:results_plotSampM}
    \vspace*{-1mm}
\end{figure}
\vspace{17.55bp}\begin{table}[!t]
    \captionsetup{font=footnotesize}
    \caption{Average MSEs in Decibels under Mismatched Priors.
    }
    \vspace{0mm}
    \label{table:missmatch}
    \footnotesize
        \centering
        \setlength{\tabcolsep}{4pt}
        {\fontsize{8}{8}\selectfont
            \begin{tabular}{cc>{\centering\arraybackslash}p{1.4cm}>{\centering\arraybackslash}p{1.4cm}>{\centering\arraybackslash}p{1.4cm}}
            \toprule
                \multirow{2}{*}[-1mm]{\shortstack{Generated\\Priors}} & \multirow{2}{*}[-1mm]{\shortstack{Assumed\\Priors}} & \multicolumn{3}{c}{Methods} \\
                \cmidrule(lr){3-5}
                && Ours (i) & Ours (ii) & Ours (iii)\\
            \midrule
            \midrule
                \multirow{3}{*}[-1mm]{SB}
                    & \multirow{1}{*}{\textbf{SB}}
                        & $\mathbf{-608.435}$ & $\mathbf{-610.582}$ & $\mathbf{-606.854}$ \\
                \cmidrule(lr){2-5}
                    & \multirow{1}{*}{SM}
                        & $-53.592$ & $-32.490$ & $-27.783$ \\
                \cmidrule(lr){2-5}
                    & \multirow{1}{*}{ST}
                        & $-17.877$ & $-16.859$ & $-16.785$ \\

            \midrule
                \multirow{3}{*}[-1mm]{SM}
                    & \multirow{1}{*}{SB}
                        & $-21.208$ & $-19.986$ & $-17.711$ \\
                \cmidrule(lr){2-5}
                    & \multirow{1}{*}{\textbf{SM}}
                        & $\mathbf{-21.281}$ & $\mathbf{-20.668}$ & $\mathbf{-20.338}$ \\
                \cmidrule(lr){2-5}
                    & \multirow{1}{*}{ST}
                        & $-16.863$ & $-16.293$ & $-16.276$ \\

            \midrule
                \multirow{3}{*}[-1mm]{ST}
                    & \multirow{1}{*}{SB}
                        & $-8.000$ & $-8.001$ & $-7.970$ \\
                \cmidrule(lr){2-5}
                    & \multirow{1}{*}{SM}
                        & $-8.001$ & $-8.000$ & $-7.981$ \\
                \cmidrule(lr){2-5}
                    & \multirow{1}{*}{\textbf{ST}}
                        & $\mathbf{-9.378}$ & $\mathbf{-9.348}$ & $\mathbf{-9.129}$ \\

            \bottomrule
            \end{tabular}
        }
        \\
        \small
        \raggedright
    \vspace*{0mm}
\end{table}
}
51--91\% of the elements of $\sampM$ are zero, and most elements of $\sampM$ take either $0$ or $1$ as shown in Fig.~\ref{fig:results_sampM}, resulting in clearly selecting or discarding signal values at each vertex for sampling while producing dense and non-local aggregation weights, which can also be seen in Fig.~\ref{fig:results_plotSampM}.

    Design (ii) shows an intermediate behavior between (i) and (iii) because it includes the squared Frobenius norm in the objective function, which tends to make each element of $\sampM$ take smaller values.
    Consequently, 51--67\% of the elements of $\sampM$ are zero as shown in Table~\ref{table:zeros}, and the other elements take a variety of values in the range $[0,1]$ as shown in Fig.~\ref{fig:results_sampM}.
    It can also be seen in Fig.~\ref{fig:results_plotSampM} that it utilizes only localized vertices compared to (i) but is not binary, unlike (iii).
    As a result, Design (ii) shows intermediate performance between (i) and (iii) in many cases as shown in Table~\ref{table:M16}.

    Table~\ref{table:runtime} shows the average runtime of each method for each graph signal type.
    As can be seen from the table, the runtime of the proposed method is longer than those of the other methods.
    This is due to the fact that it designs the sampling operator by an iterative algorithm.
    Many other methods use non-iterative
\PublicationWide{349.93bp}{%
  \begin{table*}[!t]
    \captionsetup{font=footnotesize}
    \centering
    \caption{
        Average MSEs in Decibels of the Recoveries for Monthly Average Temperature at 110 Stations in Switzerland.
    }
    \vspace{0mm}
    \label{table:swisstemp}
    \footnotesize
    \begin{minipage}{1\textwidth}
        \centering
        \setlength{\tabcolsep}{4.6pt}
        {\fontsize{8}{8}\selectfont
            \begin{tabular}{>{\centering\arraybackslash}p{0.8cm}>{\centering\arraybackslash}cccccccccc}
            \toprule
                \multirow{2}{*}[-1mm]{Priors} & \multicolumn{10}{c}{Methods} \\
                \cmidrule(lr){2-11}
                & NLPD~\cite{chen2015discrete} & SASB~\cite{hara2021design} & AVM~\cite{jayawant2022practical} & GSAO~\cite{wang2023fast} & GSSS~\cite{hara2022gsss} & SUST~\cite{hara2023graph} & RAND & \textbf{Ours (i)} & \textbf{Ours (ii)} & \textbf{Ours (iii)}\\
            \midrule
            \midrule
                \multirow{2}{*}{SB}
                        & $-57.641$ & $-26.233$ & $-34.613$ & $-56.233$ & $-61.780$ & $-59.376$ & $-16.768$ & $\mathbf{-65.530}$ & $\underline{-64.888}$ & $-62.632$\\
                        & $ (-59.298) $ & $ (-26.480) $ & $ (-36.707) $ & $ (-55.769) $ & $ (-63.697) $ & $ (-62.799) $ & $ (-10.455) $ & $ (-69.392) $ & $ (-68.919) $ & $ (-66.239) $\\
                    \multirow{2}{*}{+noise}
                        & $-27.162$ & $10.264$ & $3.880$ & $-26.530$ & $-8.672$ & $-38.280$ & $-15.617$ & $\mathbf{-63.938}$ & $\underline{-54.214}$ & $-51.027$ \\
                        & $ (-33.593) $ & $ (10.505) $ & $ (2.899) $ & $ (-30.591) $ & $ (-13.952) $ & $ (-51.803) $ & $ (-14.262) $ & $ (-69.606) $ & $ (-65.307) $ & $ (-58.186) $\\
            \midrule
                \multirow{2}{*}{SM}
                        & $-$ & $-$ & $-$ & $-$ & $-46.571$ & $-48.520$ & $-$ & $\underline{-54.716}$ & $\mathbf{-54.924}$ & $-51.131$\\
                        & $-$ & $-$ & $-$ & $-$ & $ (-50.152) $ & $ (-58.347) $ & $-$ & $ (-60.681) $ & $ (-64.570) $ & $ (-58.130) $\\
                    \multirow{2}{*}{+noise}
                        & $-$ & $-$ & $-$ & $-$ & $-43.346$ & $-43.543$ & $-$ & $\mathbf{-52.761}$ & $\underline{-51.217}$ & $-49.258$ \\
                        & $-$ & $-$ & $-$ & $-$ & $ (-45.078) $ & $ (-56.499) $ & $-$ & $ (-61.273) $ & $ (-66.295) $ & $ (-58.360) $ \\
            \midrule
                \multirow{2}{*}{ST}
                        & $-$ & $-$ & $-$ & $-$ & $-46.120$ & $-60.159$ & $-$ &  $\mathbf{-68.427}$ & $-62.378$ & $\underline{-64.674}$\\
                        & $-$ & $-$ & $-$ & $-$ & $ (-47.723) $ & $ (-64.229) $ & $-$ & $ (-72.535) $ & $ (-63.775) $ & $ (-65.684) $\\
                    \multirow{2}{*}{+noise}
                        & $-$ & $-$ & $-$ & $-$ & $-37.616$ & $-52.395$ & $-$ & $\mathbf{-64.049}$ & $-56.420$ & $\underline{-63.748}$ \\
                        & $-$ & $-$ & $-$ & $-$ & $ (-38.781) $ & $ (-54.756) $ & $-$ & $ (-65.429) $ & $ (-59.455) $ & $ (-68.509) $ \\
            \bottomrule
            \end{tabular}
        }
        \\ \vspace*{1mm}
        \small
        \raggedright
    \end{minipage}
    \vspace{0mm}
\end{table*}
\vspace{31.95bp}\def\resultA{$-58.475$}
\def\resultB{$-27.794$}
\def\resultC{$-31.997$}
\def\resultD{$-58.565$}
\def\resultE{$-61.976$}
\def\resultF{$-57.447$}
\def\resultG{$-36.310$}
\def\resultH{$\mathbf{-64.072}$}
\def\resultI{$\underline{-63.256}$}
\def\resultJ{$-62.173$}

\def\resultNoiseA{$-28.963$}
\def\resultNoiseB{$9.629$}
\def\resultNoiseC{$14.030$}
\def\resultNoiseD{$-28.887$}
\def\resultNoiseE{$-6.081$}
\def\resultNoiseF{$-37.227$}
\def\resultNoiseG{$-8.505$}
\def\resultNoiseH{$-62.344$}
\def\resultNoiseI{$-51.805$}
\def\resultNoiseJ{$-50.533$}

\begin{figure*}[t]
    \centering
    \captionsetup{font=footnotesize}
    \captionsetup[subfigure]{font=scriptsize}

    \begin{subfigure}[t]{.040\textwidth}
        \hspace*{\fill}
    \end{subfigure}
    \hfill
    \begin{subfigure}[t]{.135\textwidth}
        \includegraphics[width=\linewidth]{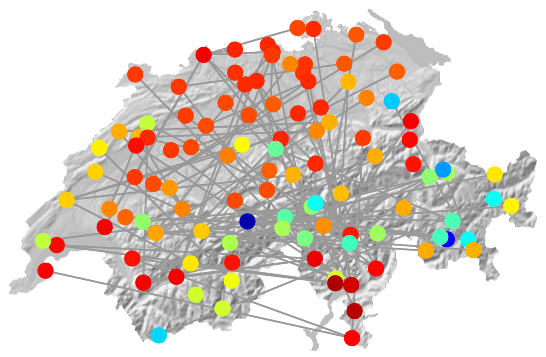}
        \vspace*{-6mm}
        \subcaption{Original \\ MSE [dB]}
    \end{subfigure}
    \hfill
    \begin{subfigure}[t]{.135\textwidth}
        \includegraphics[width=\linewidth]{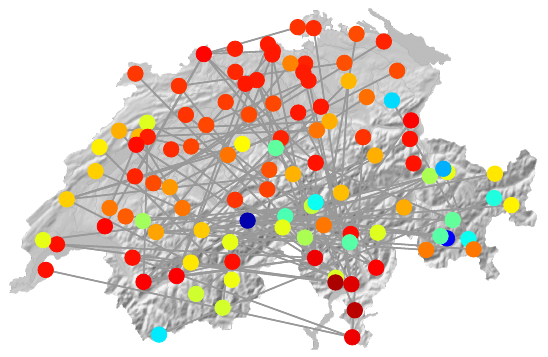}
        \vspace*{-6mm}
        \subcaption{NLPD~\cite{chen2015discrete} \\ \resultA  }
    \end{subfigure}
    \hfill
    \begin{subfigure}[t]{.135\textwidth}
        \includegraphics[width=\linewidth]{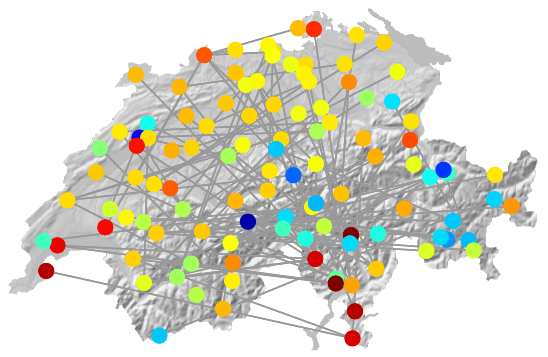}
        \vspace*{-6mm}
        \subcaption{SASB~\cite{hara2021design} \\ \resultB  }
    \end{subfigure}
    \hfill
    \begin{subfigure}[t]{.135\textwidth}
        \includegraphics[width=\linewidth]{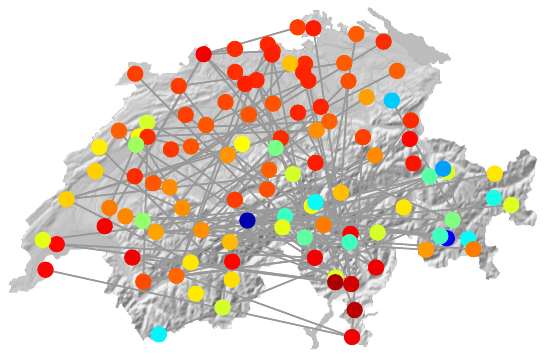}
        \vspace*{-6mm}
        \subcaption{AVM~\cite{jayawant2022practical} \\ \resultC  }
    \end{subfigure}
    \hfill
    \begin{subfigure}[t]{.135\textwidth}
        \includegraphics[width=\linewidth]{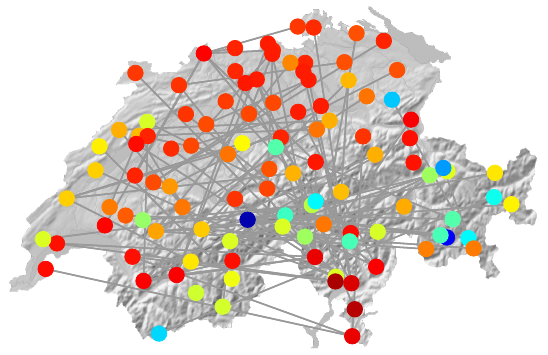}
        \vspace*{-6mm}
        \subcaption{GSAO~\cite{wang2023fast} \\ \resultD  }
    \end{subfigure}
    \hfill
    \begin{subfigure}[t]{.135\textwidth}
        \includegraphics[width=\linewidth]{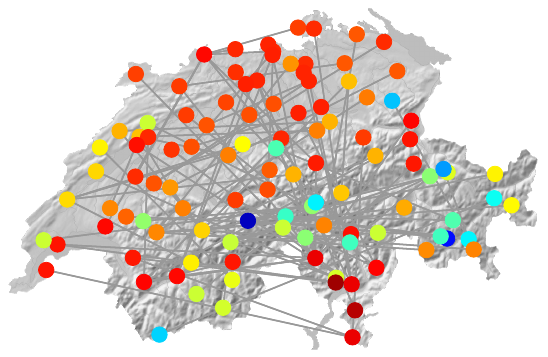}
        \vspace*{-6mm}
        \subcaption{GSSS~\cite{hara2022gsss} \\ \resultE  }
    \end{subfigure}
    \hfill
    \begin{subfigure}{.021\textwidth}
        \includegraphics[width=\linewidth]{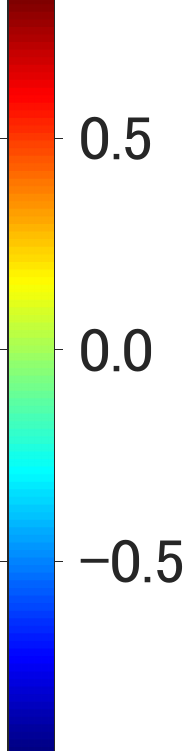}
    \end{subfigure}
    \hfill
    \begin{subfigure}[t]{.040\textwidth}
        \hspace*{\fill}
    \end{subfigure}

    \begin{subfigure}[t]{.040\textwidth}
        \hspace*{\fill}
    \end{subfigure}
    \hfill
    \begin{subfigure}[t]{.135\textwidth}
        \hspace*{\fill}
    \end{subfigure}
    \hfill
    \begin{subfigure}[t]{.135\textwidth}
        \includegraphics[width=\linewidth]{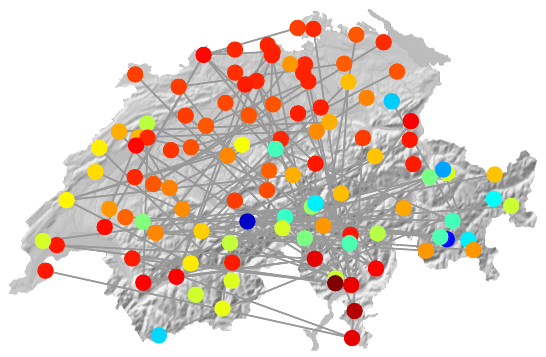}
        \vspace*{-6mm}
        \subcaption{SUST~\cite{hara2023graph} \\ \resultF }
    \end{subfigure}
    \hfill
    \begin{subfigure}[t]{.135\textwidth}
        \includegraphics[width=\linewidth]{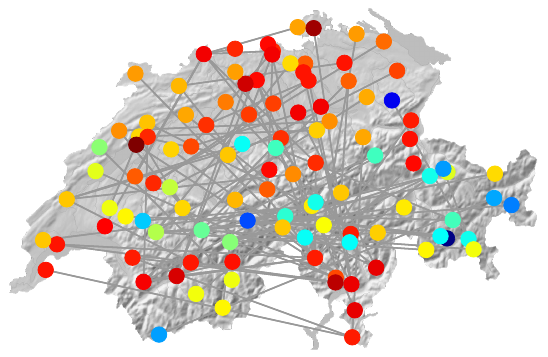}
        \vspace*{-6mm}
        \subcaption{RAND \\ \resultG }
    \end{subfigure}
    \hfill
    \begin{subfigure}[t]{.135\textwidth}
        \includegraphics[width=\linewidth]{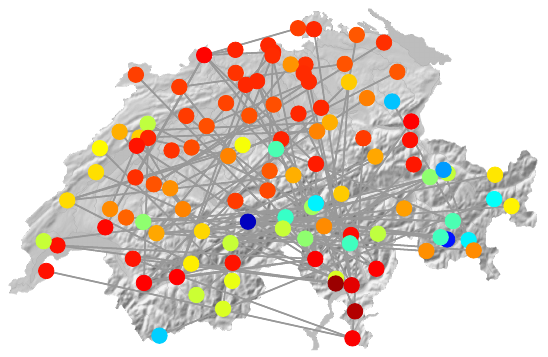}
        \vspace*{-6mm}
        \subcaption{\textbf{Ours (i)} \\ \resultH }
    \end{subfigure}
    \hfill
    \begin{subfigure}[t]{.135\textwidth}
        \includegraphics[width=\linewidth]{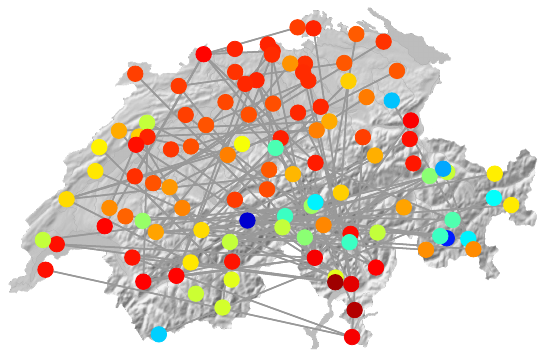}
        \vspace*{-6mm}
        \subcaption{\textbf{Ours (ii)} \\ \resultI }
    \end{subfigure}
    \hfill
    \begin{subfigure}[t]{.135\textwidth}
        \includegraphics[width=\linewidth]{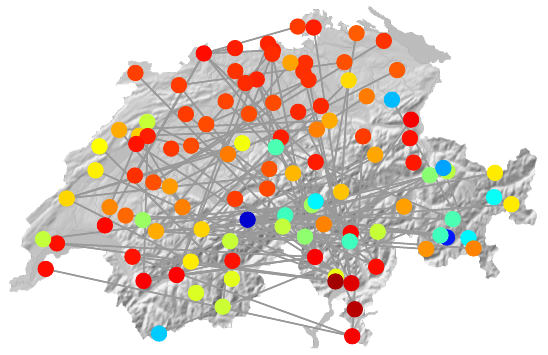}
        \vspace*{-6mm}
        \subcaption{\textbf{Ours (iii)} \\ \resultJ }
    \end{subfigure}
    \hfill
        \begin{subfigure}{.021\textwidth}
        \includegraphics[width=\linewidth]{12_fig/fig/CH/01withoutNoise/colorbar.pdf}
    \end{subfigure}
    \hfill
    \begin{subfigure}[t]{.040\textwidth}
        \hspace*{\fill}
    \end{subfigure}

    \vspace*{-1mm}
    \begin{minipage}[t]{384.3bp}
    \caption{
        Signal recovery experiments for the average temperature of Switzerland in March under the subspace prior.
    }
\label{fig:swisstemp}
    \end{minipage}
\vspace{0mm}
\end{figure*}
}
algorithms, such as greedy methods, for designing the sampling operator, resulting in shorter runtimes.
    Even though the runtime including designing the sampling operator is relatively high, the recovery accuracy is superior in most cases.
    \blue{Note that the design of the sampling operator, together with the computation of the corresponding correction and reconstruction operators, is an offline process performed only once.
    Once they are obtained, the sampling and recovery of each graph signal can be performed quickly with only matrix-vector products, since the recovered graph signal is given by $\grecsig = \recM \corM \sampMT \gsig$ as shown in Eq.~\eqref{gsig_recovery}.}

\subsection{Prior Mismatch Scenario}
    We also conducted experiments to evaluate the performance of our proposed method in scenarios where there is a mismatch between the assumed prior used for designing the sampling operator and the actual prior used to generate the graph signals.
    We generated BL signals for the subspace prior, GMRF signals for the smoothness prior, and SGS signals for the stochastic prior with the same setup as in the synthetic graph signal experiments.
    The parameters of the experiments and the proposed methods are the same as those in the synthetic experiment including the number of vertices on a graph $\numv = 256$ and the size of the sampled signal $\nums = 16$.
    We assumed the unconstrained case for the reconstruction operator.
    We ran 20 independent trials and compared the results using the averaged MSE between the original and recovered signals.
    The results are summarized in Table~\ref{table:missmatch}.

    \PublicationColumnTwelve
    As can be seen from the table, the recovery performance de-\linebreak
creases when there is a mismatch. In particular, when assuming\linebreak
priors different from the subspace prior for signals generated\linebreak
with the subspace prior, the performance significantly degrades,\linebreak
and it is confirmed that the performance also degrades for the\linebreak
other priors. From this, it is shown that sampling and recovery\linebreak
assuming appropriate priors are important for signals generated\linebreak
with any prior.

\subsection{Real-World Data}
\subsubsection{Setup}
    \looseness=-1 We have also conducted sampling and recovery experiments using real-world data.
    We used temperature data from $\numv = 110$ stations in Switzerland from 2015 to 2024~\cite{MeteoSwissData}, and normalized the temperatures, originally in degrees Celsius, so that the maximum absolute value of the average monthly temperature is $1$.
    The size of the sampled signal was set to $\nums = 28$.
    We used the data from the first five years to estimate the graph Laplacian of each month using the graph learning method~\cite{dong2016learning}.
    As for the generator matrix for the subspace prior, we used the first $K = 28$ columns of the unitary matrix $\mathbf{U}$ obtained by the eigendecomposition of the estimated graph Laplacian.
    As for the smoothness prior, we used the same spectral response $\smopefil$ of the smoothness operator $\smopeM$ as in the synthetic data experiment.
    As for the known covariance matrix for the stochastic prior, we estimated it from the first five years' data with the method proposed in~\cite{perraudin2017stationary}.
    Then we performed experiments with the average monthly temperature from the remaining five
\PublicationPageThirteen
years' data.
    The same parameter settings as in the synthetic data experiment were used, and we assumed the unconstrained case for the reconstruction operator.
    We compared the averaged MSE for each prior.

\subsubsection{Results}
    The experimental results are shown in Table~\ref{table:swisstemp} and Fig.~\ref{fig:swisstemp}.
    As in Table~\ref{table:M16}, the numbers in parentheses represent the standard deviations in decibels.
    Since all the rows in Table~\ref{table:swisstemp} correspond to the same data with different assumed priors, the results of NLPD, AVM, GSAO, and RAND, which do not distinguish between the priors, and those of SASB, which is applicable only to the subspace prior, are shown only in the SB rows.
    The table shows that the recovery results by the proposed method are superior.
    This result from real-world data also indicates that applying DC optimization for sampling to achieve the best possible recovery of beyond bandlimited graph signals is effective.

\vspace{-15.8bp}
\section{Conclusion}
\label{sec:conclusion}
\looseness=-1 In this paper, we addressed the challenge of achieving the best possible recovery for sampling beyond bandlimited graph signals under arbitrary priors in the vertex domain by applying DC optimization.
We formulated the design of an aggregation sampling operator as a problem with the full rank constraint to achieve the best possible recovery based on the generalized sampling theory.
To handle the constraint, we transformed the problem into a DC optimization problem by relaxing the constraint using the nuclear norm.
To solve the problem, we developed a solver based on the GDPGDC algorithm, which ensures convergence to a critical point.
To evaluate the effectiveness of our method, we performed sampling and recovery experiments on various types of graph signals, comparing the MSE between the original and recovered signals.
The results demonstrated the effectiveness of our approach and highlighted its value in sampling graph signals under arbitrary priors in the vertex domain through achieving the best possible recovery via DC optimization.
\blue{An important direction for future work is to reduce the computational complexity for designing the sampling operator, especially to improve the scalability of the proposed method to larger graphs.}

\vspace{-12bp}
\appendices
\section{Derivation of Eq.~\texorpdfstring{\eqref{eq:proj_FNandBox}}{(41)}}
\label{appendix:calc_FNandBox}
Due to~\eqref{eq:prox_f2}, the proximity operator of $\gamma_1(\fsamp + \iota_{\sampMSet})$ for Design (ii) can be transformed as follows:
\begin{align}
    \label{eq:proof_FNandBox_1}
    &\prox_{\gamma_1(\paramBox \FN{\cdot}^2 + \indi{\BoxconNM}{\cdot})} (\mat) \nonumber \\
    &= \scalebox{0.95}{$\displaystyle\argmin_{\matt} \paramBox \FN{\matt}^{2} + \indi{\BoxconNM}{\matt} + \frac{1}{2\gamma_1} \FN{\mat - \matt}^{2}$} \nonumber \\
    &= \scalebox{0.95}{$\displaystyle\argmin_{\matt} \sum_{i,j} \left[ \paramBox \elee^{2} + \indi{\Boxcon}{\elee} + \frac{1}{2\gamma_1} \left(\ele - \elee \right)^{2} \right]$}.
\end{align}
Since there are no interaction terms between different $i$, $j$, the optimization can be performed independently for each element.
From properties of the convex optimization of single-variable functions with box constraints, we have
\begin{align}
    \label{eq:proof_FNandBox_3}
    &\argmin_{\elee} \left[ \paramBox \elee^{2} + \indi{\Boxcon}{\elee} + \frac{1}{2\gamma_1} (\ele - \elee)^{2} \right] \nonumber \\
    &=
    \begin{cases}
        \lbound, \ &\mbox{if} \ \prox_{\gamma_1 \paramBox \FN{\cdot}^2} (\ele) < \lbound; \\
        \prox_{\gamma_1 \paramBox \FN{\cdot}^2} (\ele), \ &\mbox{if} \ \lbound \leq \prox_{\gamma_1 \paramBox \FN{\cdot}^2} (\ele) \leq \ubound; \\
        \ubound, \ &\mbox{if} \ \prox_{\gamma_1 \paramBox \FN{\cdot}^2} (\ele) > \ubound
    \end{cases} \nonumber \\
    &= \max \left\{ \lbound, \min \left\{ \prox_{\gamma_1 \paramBox \FN{\cdot}^2} (\ele), \ubound \right\} \right\},
\end{align}
where $\prox_{\gamma_1 \paramBox \FN{\cdot}^2} (\ele) = {\ele}/(1 + 2 \gamma_1 \paramBox)$~\cite{bauschke2017correction}.
Therefore, $\prox_{\gamma_1( \paramBox \FN{\cdot}^2 + \indi{\BoxconNM}{\cdot})} (\mat)$ is calculated as follows: for all $i$ from $1$ to $\numv$ and $j$ from $1$ to $\nums$,
\begin{align}
    \label{eq:proof_FNandBox_4}
    &\left[\prox_{\gamma_1( \paramBox \FN{\cdot}^2 + \indi{\BoxconNM}{\cdot})} (\mat)\right]_{ij} \nonumber \\
    &= \max \left\{ \lbound, \min \left\{ \prox_{\gamma_1 \paramBox \FN{\cdot}^2} (\ele), \ubound \right\} \right\}.
\end{align}

\section{Derivation of Eq.~\texorpdfstring{\eqref{eq:proj_L1andBox}}{(42)}}
\label{appendix:calc_L1andBox}
Due to~\eqref{eq:prox_f2}, the proximity operator of $\gamma_1(\fsamp + \iota_{\sampMSet})$ for Design (iii) can be transformed as follows:
\begin{align}
    \label{eq:proof_L1andBox_1}
    &\prox_{\gamma_1(\paramBox \Lone{\cdot} + \indi{\BoxconNM}{\cdot})} (\mat) \nonumber \\
    &=\scalebox{0.95}{$\displaystyle\argmin_{\matt} \paramBox \Lone{\matt} + \indi{\BoxconNM}{\matt} + \frac{1}{2\gamma_1} \FN{\mat - \matt}^{2}$} \nonumber \\
    &=\scalebox{0.95}{$\displaystyle\argmin_{\matt} \sum_{i,j} \left[ \paramBox | \elee | + \indi{\Boxcon}{\elee} + \frac{1}{2\gamma_1} \left(\ele - \elee \right)^{2} \right]$}.
\end{align}
Since there are no interaction terms between different $i$, $j$, the optimization can be performed independently for each element.
From properties of the convex optimization of single-variable functions with box constraints, we have
\begin{align}
    \label{eq:proof_L1andBox_2}
    &\argmin_{\elee} \left[ \paramBox | \elee | + \indi{\Boxcon}{\elee} + \frac{1}{2\gamma_1} \left(\ele - \elee \right)^{2} \right] \nonumber \\
    &= \begin{cases}
        \lbound,
            &\mathrm{if} \ \prox_{\gamma_1 \paramBox \Lone{\cdot}}\left(\ele\right) < \lbound; \\
        \prox_{\gamma_1 \paramBox \Lone{\cdot}}\left(\ele\right),
            &\mathrm{if} \ \lbound \leq \prox_{\gamma_1 \paramBox \Lone{\cdot}}\left(\ele\right) \leq \ubound; \\
        \ubound,
            &\mathrm{if} \ \prox_{\gamma_1 \paramBox \Lone{\cdot}}\left(\ele\right) > \ubound
    \end{cases} \nonumber \\
    &= \max \left\{ \lbound, \min\left\{ \prox_{\gamma_1 \paramBox \Lone{\cdot}}\left(\ele\right), \ubound \right\}\right\},
\end{align}
where $\prox_{\gamma_1 \paramBox \Lone{\cdot}}\left(\ele\right)$ is given by~\cite{bauschke2017correction}:
\begin{align}
\label{eq:proof_L1andBox_3}
    &\prox_{\gamma_1 \paramBox \Lone{\cdot}}\left(\ele\right) \nonumber\\
    &=
    \begin{cases}
        \ele + \gamma_1 \paramBox, \ &\mbox{if} \ \ele < - \gamma_1 \paramBox; \\
        0, \ &\mbox{if} \ -\gamma_1 \paramBox \leq \ele \leq \gamma_1 \paramBox; \\
        \ele - \gamma_1 \paramBox, \ &\mbox{if} \ \ele > \gamma_1 \paramBox.
    \end{cases}
\end{align}
Therefore, $\prox_{\gamma_1(\paramBox \Lone{\cdot} + \indi{\BoxconNM}{\cdot})} (\mat)$ is calculated as follows: for all $i$ from $1$ to $\numv$ and $j$ from $1$ to $\nums$,
\begin{align}
    &\left[\prox_{\gamma_1( \paramBox \Lone{\cdot} + \indi{\BoxconNM}{\cdot})} (\mat)\right]_{ij} \nonumber \\
    &= \max \left\{ \lbound, \min\left\{ \prox_{\gamma_1 \paramBox \Lone{\cdot}}\left(\ele\right), \ubound \right\}\right\}.
\end{align}

\PublicationFlushDisplayColumn{6.476bp}\clearpage
\bibliographystyle{IEEEtran}
\bibliography{09_refs}

@string{ICASSP="Proc. IEEE Int. Conf. Acoust., Speech, Signal Process. (ICASSP)"}

@string{EUSIPCO="Proc. Eur. Signal Process. Conf. (EUSIPCO)"}

@article{shuman2013emerging,
  month=may,
  author={Shuman, D. I. and Narang, S. K. and Frossard, P. and Ortega, A. and Vandergheynst, P.},
  journal={IEEE Signal Process. Mag.},
  title={The emerging field of signal processing on graphs: Extending high-dimensional data analysis to networks and other irregular domains},
  year={2013},
  volume={30},
  number={3},
  pages={83-98}
}

@article{sandryhaila2013discrete,
  month=apr,
  title={Discrete signal processing on graphs},
  author={Sandryhaila, A. and Moura, J. M. F.},
  journal={IEEE Trans. Signal Process.},
  volume={61},
  number={7},
  pages={1644--1656},
  year={2013},

}

@article{ortega2018graph,
  month=may,
  author={Ortega, A. and Frossard, P. and Kova{\v{c}}evi{\'c}, J. and Moura, J. M. F. and Vandergheynst, P.},
  journal={Proc. IEEE},
  title={Graph Signal Processing: Overview, Challenges, and Applications},
  year={2018},
  volume={106},
  number={5},
  pages={808-828}
}

@article{chen2015discrete,
  month=dec,
  author={Chen, S. and Varma, R. and Sandryhaila, A. and Kova{\v{c}}evi{\'c}, J.},
  journal={IEEE Trans. Signal Process.},
  title={Discrete Signal Processing on Graphs: Sampling Theory},
  year={2015},
  volume={63},
  number={24},
  pages={6510-6523}
}

@article{tanaka2020sampling,
  month=nov,
  author={Tanaka, Y. and Eldar, Y. C. and Ortega, A. and Cheung, G.},
  journal={IEEE Signal Process. Mag.},
  title={Sampling Signals on Graphs: From Theory to Applications},
  year={2020},
  volume={37},
  number={6},
  pages={14-30}
}

@article{marques2015sampling,
  month=apr,
  title={Sampling of graph signals with successive local aggregations},
  author={Marques, A. G. and Segarra, S. and Leus, G. and Ribeiro, A.},
  journal={IEEE Trans. Signal Process.},
  volume={64},
  number={7},
  pages={1832--1843},
  year={2016},
}

@article{anis2016efficient,
  month=jul,
  author={Anis, A. and Gadde, A. and Ortega, A.},
  journal={IEEE Trans. Signal Process.},
  title={Efficient Sampling Set Selection for Bandlimited Graph Signals Using Graph Spectral Proxies},
  year={2016},
  volume={64},
  number={14},
  pages={3775-3789}
}

@article{tsitsvero2016signals,
  month=sep,
  title={Signals on graphs: Uncertainty principle and sampling},
  author={Tsitsvero, M. and Barbarossa, S. and Di Lorenzo, P.},
  journal={IEEE Trans. Signal Process.},
  volume={64},
  number={18},
  pages={4845--4860},
  year={2016},

}

@article{valsesia2018sampling,
  month=jun,
  title={Sampling of graph signals via randomized local aggregations},
  author={Valsesia, D. and Fracastoro, G. and Magli, E.},
  journal={IEEE Trans. Signal Inf. Process. Netw.},
  volume={5},
  number={2},
  pages={348--359},
  year={2019},

}

@article{tanaka2018spectral,
  month=jul,
  title={Spectral domain sampling of graph signals},
  author={Tanaka, Y.},
  journal={IEEE Trans. Signal Process.},
  volume={66},
  number={14},
  pages={3752--3767},
  year={2018},

}

@article{sakiyama2019eigendecomposition,
  month=may,
  author={Sakiyama, A. and Tanaka, Y. and Tanaka, T. and Ortega, A.},
  journal={IEEE Trans. Signal Process.},
  title={Eigen\-de\-com\-po\-si\-tion-Free Sampling Set Selection for Graph Signals},
  year={2019},
  volume={67},
  number={10},
  pages={2679-2692}
}

@article{bai2020fast,
  title={Fast graph sampling set selection using {G}ershgorin disc alignment},
  author={Bai, Y. and Wang, F. and Cheung, G. and Nakatsukasa, Y. and Gao, W.},
  journal={IEEE Trans. Signal Process.},
  volume={68},
  pages={2419--2434},
  year={2020},

}

@article{jayawant2022practical,
  title={Practical graph signal sampling with log-linear size scaling},
  author={Jayawant, A. and Ortega, A.},
  journal={Signal Process.},
  volume={194},
  note={{Art. no. 108436}},
  year={2022}
}

@article{elder2009beyond,
  month=may,
  author={Eldar, Y. C. and Michaeli, T.},
  journal={IEEE Signal Process. Mag.},
  title={Beyond bandlimited sampling},
  year={2009},
  volume={26},
  number={3},
  pages={48-68}
}

@book{eldar2015sampling,
  title={Sampling Theory: Beyond Bandlimited Systems},
  author={Eldar, Y. C.},
  year={2015},
  publisher={Cambridge Univ. Press},
  address={Cambridge, U.K.}
}

@inproceedings{chepuri2018graphsamp,
  author={Chepuri, S. P. and Eldar, Y. C. and Leus, G.},
  booktitle={Proc. IEEE Int. Conf. Acoust., Speech, Signal Process. (ICASSP)},
  title={Graph Sampling with and Without Input Priors},
  year={2018},
  volume={},
  number={},
  pages={4564-4568}
}

@article{tanaka2020gensamp,
  author={Tanaka, Y. and Eldar, Y. C.},
  journal={IEEE Trans. Signal Process.},
  title={Generalized Sampling on Graphs With Subspace and Smoothness Priors},
  year={2020},
  volume={68},
  number={},
  pages={2272-2286}
}

@article{hara2023graph,
  title={Graph signal sampling under stochastic priors},
  author={Hara, J. and Tanaka, Y. and Eldar, Y. C.},
  journal={IEEE Trans. Signal Process.},
  year={2023},
  volume={71},
  number={},
  pages={1421-1434},

}

@inproceedings{hara2022gsss,
  author={Hara, J. and Tanaka, Y.},
  booktitle={Proc. IEEE Int. Conf. Acoust., Speech, Signal Process. (ICASSP)},
  title={Sampling Set Selection for Graph Signals under Arbitrary Signal Priors},
  year={2022},
  volume={},
  number={},
  pages={5732-5736}
}

@inproceedings{hara2021design,
  title={Design of graph signal sampling matrices for arbitrary signal subspaces},
  author={Hara, J. and Yamada, K. and Ono, S. and Tanaka, Y.},
  booktitle={Proc. IEEE Int. Conf. Acoust., Speech, Signal Process. (ICASSP)},
  pages={5275--5279},
  year={2021}
}

@article{banert2019general,
  title={A general double-proximal gradient algorithm for {DC} programming},
  author={Banert, S. and Bo{\c{t}}, R. I.},
  journal={Math. Program.},
  volume={178},
  number={1-2},
  pages={301--326},
  year={2019},
  publisher={Springer}
}

@inproceedings{chen2016representations,
  title={Representations of piecewise smooth signals on graphs},
  author={Chen, S. and Varma, R. and Singh, A. and Kova{\v{c}}evi{\'c}, J.},
  booktitle={Proc. IEEE Int. Conf. Acoust., Speech, Signal Process. (ICASSP)},
  pages={6370--6374},
  year={2016},
}

@phdthesis{fazel2002matrix,
  title={Matrix rank minimization with applications},
  author={Fazel, M.},
  year={2002},
  school={Stanford Univ.}
}

@book{bauschke2017correction,
  author={Bauschke, H. H. and Combettes, P. L.},
  title={Convex Analysis and Monotone Operator Theory in {H}ilbert Spaces},
  edition={2nd},
  publisher={Springer},
  address={Cham, Switzerland},
  year={2017}
}

@misc{perraudin2014gspbox,
  title={{GSPBOX}: A toolbox for signal processing on graphs},
  author={Perraudin, N. and Paratte, J. and Shuman, D. I. and Martin, L. and Kalofolias, V. and Vandergheynst, P. and Hammond, D. K.},
  year={2014},
  note = {[Online]. Available: https://arxiv.org/abs/1408.5781},
}

@article{egilmez2018graph,
  month=jun,
  title={Graph learning from filtered signals: Graph system and diffusion kernel identification},
  author={Egilmez, H. E. and Pavez, E. and Ortega, A.},
  journal={IEEE Trans. Signal Inf. Process. Netw.},
  volume={5},
  number={2},
  pages={360--374},
  year={2019},

}

@article{dong2019learning,
  month=may,
  title={Learning graphs from data: A signal representation perspective},
  author={Dong, X. and Thanou, D. and Rabbat, M. and Frossard, P.},
  journal={IEEE Signal Process. Mag.},
  volume={36},
  number={3},
  pages={44--63},
  year={2019},

}

@inproceedings{ono2015total,
  title={Total generalized variation for graph signals},
  author={Ono, S. and Yamada, I. and Kumazawa, I.},
  booktitle={Proc. IEEE Int. Conf. Acoust., Speech, Signal Process. (ICASSP)},
  pages={5456--5460},
  year={2015},
}

@article{onuki2016graph,
  month=jun,
  title={Graph signal denoising via trilateral filter on graph spectral domain},
  author={Onuki, M. and Ono, S. and Yamagishi, M. and Tanaka, Y.},
  journal={IEEE Trans. Signal Inf. Process. Netw.},
  volume={2},
  number={2},
  pages={137--148},
  year={2016},

}

@article{nagahama2022graph,
  title={Graph signal restoration using nested deep algorithm unrolling},
  author={Nagahama, M. and Yamada, K. and Tanaka, Y. and Chan, S. H. and Eldar, Y. C.},
  journal={IEEE Trans. Signal Process.},
  volume={70},
  pages={3296--3311},
  year={2022},

}

@article{cheung2018graph,
  month=may,
  title={Graph spectral image processing},
  author={Cheung, G. and Magli, E. and Tanaka, Y. and Ng, M. K.},
  journal={Proc. IEEE},
  volume={106},
  number={5},
  pages={907--930},
  year={2018},

}

@article{wang2019dynamic,
  title={Dynamic graph {CNN} for learning on point clouds},
  author={Wang, Y. and Sun, Y. and Liu, Z. and Sarma, S. E. and Bronstein, M. M. and Solomon, J. M.},
  journal={ACM Trans. Graph.},
  volume={38},
  number={5},
  pages={1--12},
  year={2019},
  publisher={Acm New York, NY, USA}
}

@article{bronstein2017geometric,
  month=jul,
  title={Geometric deep learning: going beyond {E}uclidean data},
  author={Bronstein, M. M. and Bruna, J. and LeCun, Y. and Szlam, A. and Vandergheynst, P.},
  journal={IEEE Signal Process. Mag.},
  volume={34},
  number={4},
  pages={18--42},
  year={2017},

}

@article{agaskar2013spectral,
  month=jul,
    author={Agaskar, A. and Lu, Y. M.},
    journal={IEEE Trans. Inf. Theory},
    title={A Spectral Graph Uncertainty Principle},
    year={2013},
    volume={59},
    number={7},
    pages={4338-4356},
    doi={10.1109/TIT.2013.2252233}
}

@article{sandryhaila2014big,
  month=sep,
  title={Big data analysis with signal processing on graphs: Representation and processing of massive data sets with irregular structure},
  author={Sandryhaila, A. and Moura, J. M. F.},
  journal={IEEE Signal Process. Mag.},
  volume={31},
  number={5},
  pages={80--90},
  year={2014},

}

@article{narang2012perfect,
  month=jun,
  title={Perfect reconstruction two-channel wavelet filter banks for graph structured data},
  author={Narang, S. K. and Ortega, A.},
  journal={IEEE Trans. Signal Process.},
  volume={60},
  number={6},
  pages={2786--2799},
  year={2012},

}

@article{shuman2015multiscale,
  month=apr,
  title={A multiscale pyramid transform for graph signals},
  author={Shuman, D. I. and Faraji, M. J. and Vandergheynst, P.},
  journal={IEEE Trans. Signal Process.},
  volume={64},
  number={8},
  pages={2119--2134},
  year={2016},

}

@article{chen2014semi,
  month=jun,
  title={Semi-supervised multiresolution classification using adaptive graph filtering with application to indirect bridge structural health monitoring},
  author={Chen, S. and Cerda, F. and Rizzo, P. and Bielak, J. and Garrett, J. H. and Kova{\v{c}}evi{\'c}, J.},
  journal={IEEE Trans. Signal Process.},
  volume={62},
  number={11},
  pages={2879--2893},
  year={2014},

}

@inproceedings{gadde2014active,
  title={Active semi-supervised learning using sampling theory for graph signals},
  author={Gadde, A. and Anis, A. and Ortega, A.},
  booktitle={Proc. 20th ACM SIGKDD Int. Conf. Knowl. Discov. Data Mining},
  pages={492--501},
  year={2014}
}

@inproceedings{perozzi2014deepwalk,
  title={{DeepWalk}: Online learning of social representations},
  author={Perozzi, B. and Al-Rfou, R. and Skiena, S.},
  booktitle={Proc. 20th ACM SIGKDD Int. Conf. Knowl. Discov. Data Mining},
  pages={701--710},
  year={2014}
}

@inproceedings{hasanzadeh2019piecewise,
  title={Piecewise stationary modeling of random processes over graphs with an application to traffic prediction},
  author={Hasanzadeh, A. and Liu, X. and Duffield, N. and Narayanan, K. R.},
  booktitle={Proc. IEEE Int. Conf. Big Data},
  pages={3779--3788},
  year={2019},
}

@article{eldar2003sampling,
  title={Sampling with arbitrary sampling and reconstruction spaces and oblique dual frame vectors},
  author={Eldar, Y. C.},
  journal={J. Fourier Anal. Appl.},
  volume={9},
  number={1},
  pages={77--96},
  year={2003},
  publisher={Springer}
}

@inproceedings{yamashita2024apsipa,
    author={Yamashita, K. and Naganuma, K. and Ono, S.},
    booktitle={Proc. Asia-Pac. Signal Inf. Process. Assoc. Annu. Summit Conf. (APSIPA ASC)},
    title={Generalized Graph Signal Sampling under Subspace Priors by Difference-of-Convex Minimization},
    year={2024},
    volume={},
    number={},
    pages={1-6},
    doi={10.1109/APSIPAASC63619.2025.10849132}
}

@inproceedings{yamashita2025controlling,
  title={Controlling the Number of Sample-Contributive Vertices in Generalized Sampling of Graph Signals},
  author={Yamashita, K. and Naganuma, K. and Ono, S.},
  booktitle={Proc. IEEE Int. Conf. Acoust., Speech, Signal Process. (ICASSP)},
  pages={1--5},
  year={2025}
}

@article{li2025sensor,
  title={Sensor placement method for water distribution networks based on sampling of non-bandlimited graph signals},
  author={Li, J. and Cai, B.},
  journal={Digit. Signal Process.},
  volume={156},
  note={{Art. no. 104809}},
  year={2025},
  publisher={Elsevier}
}

@article{dong2016learning,
  month=dec,
  title={Learning {L}aplacian matrix in smooth graph signal representations},
  author={Dong, X. and Thanou, D. and Frossard, P. and Vandergheynst, P.},
  journal={IEEE Trans. Signal Process.},
  volume={64},
  number={23},
  pages={6160--6173},
  year={2016},

}

@article{yamagata2025robust,
  title={Robust time-varying graph signal recovery for dynamic physical sensor network data},
  author={Yamagata, E. and Naganuma, K. and Ono, S.},
  journal={IEEE Trans. Signal Inf. Process. Netw.},
  year={2025},
  volume={11},
  number={},
  pages={59-70},
}

@inproceedings{hara2024sensor,
  title={Sensor Placement Problem on Networks for Sensors with Multiple Specifications},
  author={Hara, J. and Ono, S. and Higashi, H. and Tanaka, Y.},
  booktitle={Proc. Eur. Signal Process. Conf. (EUSIPCO)},
  pages={2327--2331},
  year={2024},
}

@article{wang2023fast,
  title={Fast {MSE}-based sampling of bandlimited graph signals via low-pass impulse responses},
  author={Wang, F. and Cheung, G. and Ye, M. and Li, T. and Feng, Y.-T.},
  journal={IEEE Trans. Signal Process.},
  volume={71},
  pages={4207--4223},
  year={2023},
}

@article{puy2018random,
  title={Random sampling of bandlimited signals on graphs},
  author={Puy, Gilles and Tremblay, Nicolas and Gribonval, R{\'e}mi and Vandergheynst, Pierre},
  journal={Appl. Comput. Harmon. Anal.},
  volume={44},
  number={2},
  pages={446--475},
  year={2018},
  publisher={Elsevier}
}

@misc{MeteoSwissData,
  author={{Federal Office of Meteorology and Climatology}},
  title={Homogeneous climate station data},
  year={2025},
  note={{Accessed: Dec. 19, 2025}},
  url={https://opendatadocs.meteoswiss.ch/c-climate-data/c1-climate-stations_homogeneous}
}

@article{perraudin2017stationary,
  month=jul,
  title={Stationary signal processing on graphs},
  author={Perraudin, N. and Vandergheynst, P.},
  journal={IEEE Trans. Signal Process.},
  volume={65},
  number={13},
  pages={3462--3477},
  year={2017},
}

@article{yamashita2026sampling,
  month=feb,
  title={Sampling Method for Generalized Graph Signals with Pre-selected Vertices via {DC} Optimization},
  author={Yamashita, K. and Naganuma, K. and Ono, S.},
  journal={IEEE Open J. Signal Process.},
  year={2026},
  volume={7},
  number={},
  pages={314-323},
  publisher={IEEE}
}

@article{dabush2026efficient,
  title={Efficient sampling allocation strategies for general graph-filter-based signal recovery},
  author={Dabush, L. and Routtenberg, T.},
  journal={IEEE Trans. Signal Inf. Process. Netw.},
  year={2026},
  volume={12},
  number={},
  pages={702-719},
  publisher={IEEE}
}

@article{sheng2025subset,
  title={Subset Random Sampling and Reconstruction of Finite Time-Vertex Graph Signals},
  author={Sheng, H. and Shu, Q. and Feng, H. and Hu, B.},
  journal={IEEE Trans. Signal Inf. Process. Netw.},
  year={2025},
  volume={11},
  number={},
  pages={1015-1029},
  publisher={IEEE}
}

@inproceedings{saboksayr2021learning,
    author={Saboksayr, S. S. and Mateos, G. and Cetin, M.},
    booktitle={Proc. Eur. Signal Process. Conf. (EUSIPCO)},
    title={Online Graph Learning under Smoothness Priors},
    year={2021},
    volume={},
    number={},
    pages={1820-1824}
}

@article{takemoto2022graph,
    title={Graph Spatio-Spectral Total Variation Model for Hyperspectral Image Denoising},
    author={Takemoto, S. and Naganuma, K. and Ono, S.},
    journal={IEEE Geosci. Remote Sens. Lett.},
    volume={19},
    pages={1--5},
    year={2022},

}

@inproceedings{gadde2015probabilistic,
    author={Gadde, A. and Ortega, A.},
    booktitle={Proc. IEEE Int. Conf. Acoust., Speech, Signal Process. (ICASSP)},
    title={A probabilistic interpretation of sampling theory of graph signals},
    year={2015},
    volume={},
    number={},
    pages={3257-3261}
}

@article{bolte2014proximal,
  title={Proximal alternating linearized minimization for nonconvex and nonsmooth problems},
  author={Bolte, J. and Sabach, S. and Teboulle, M.},
  journal={Math. Program.},
  volume={146},
  number={1--2},
  pages={459--494},
  year={2014}
}

@IEEEtranBSTCTL{arxiv:bstcontrol,
  CTLuse_forced_etal = {yes},
  CTLmax_names_forced_etal = {6},
  CTLnames_show_etal = {1}
}

\ifCLASSOPTIONcaptionsoff
  \newpage
\fi

\makeatletter
\def\@IEEEBIOskipN{0pt}
\def\@IEEEBIOhangwidth{1.165in}
\makeatother
\vspace{27.178bp}
\noindent\begin{minipage}[t]{\columnwidth}
\begin{IEEEbiography}[{\includegraphics[width=1in,height=1.25in,clip,keepaspectratio]{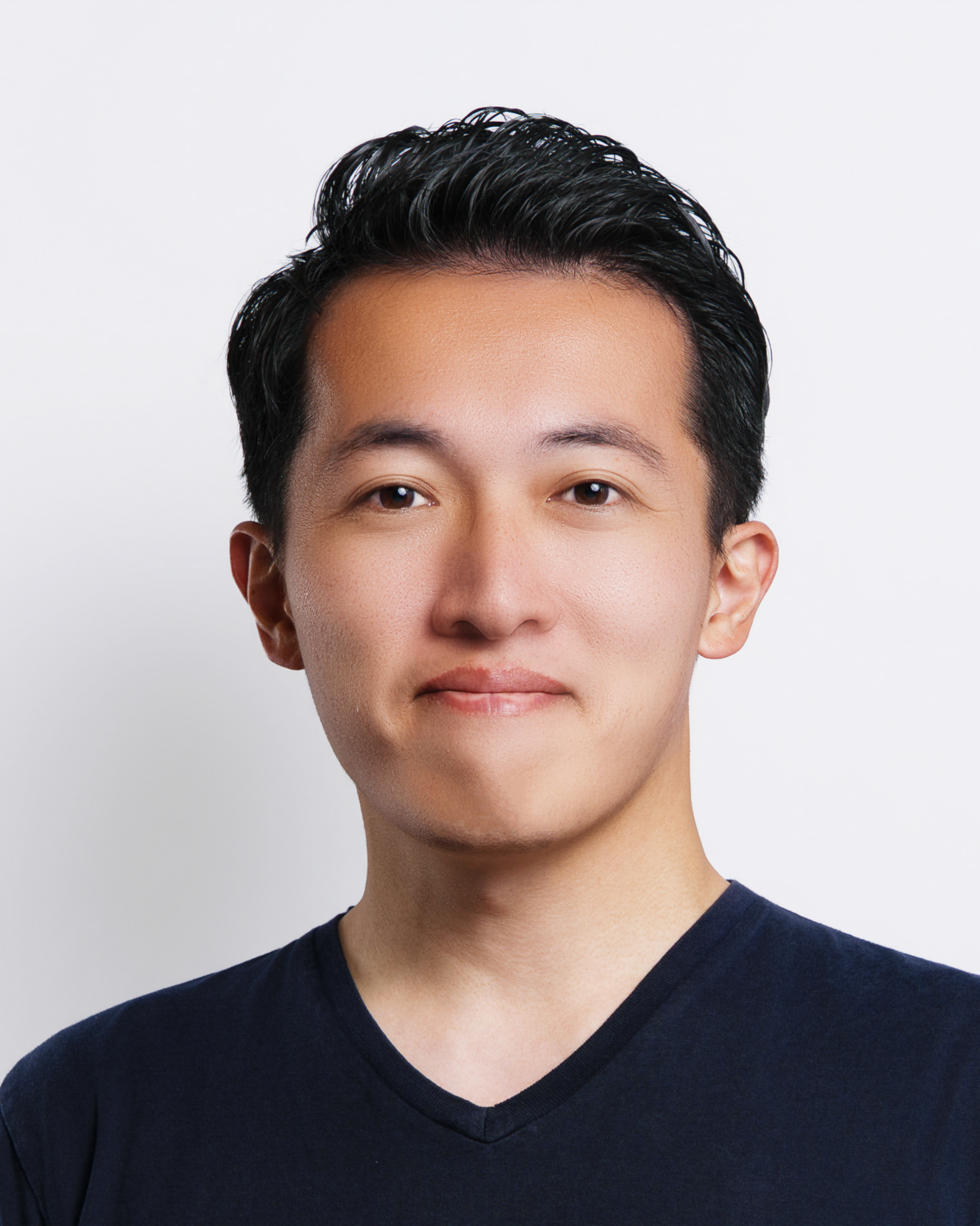}}]{Keitaro Yamashita}
    (Member, IEEE) received the B.E. degree in control engineering and the M.E. degree in decision science and technology from Tokyo Institute of Technology, Japan, in 2015 and 2018, respectively, and the Ph.D. degree in computer science from the Institute of Science Tokyo, Japan, in 2026.
    After receiving the M.E. degree, he worked for the Ministry of Economy, Trade and Industry (METI), the Government of Japan, and was in charge of policy planning for research and development, deep-tech startups, carbon neutrality, and data free flow with trust (DFFT).
    He has also worked for the Agency for Cultural Affairs, the Government of Japan, where he was in charge of the copyright policy.
    He is currently a Researcher with the Department of Computer Science, Institute of Science Tokyo.
    His research interests include graph signal processing and optimization.
    He received the Best Paper Award at the Asia-Pacific Signal and Information Processing Association Annual Summit and Conference (APSIPA ASC) in 2024 and the IEEE SPS Japan Student Conference Paper Award in 2025.
\end{IEEEbiography}
\end{minipage}\par
\newpage
\vspace*{-5.502bp}
\noindent\begin{minipage}[t]{\columnwidth}
\begin{IEEEbiography}[{\includegraphics[width=1in,height=1.25in,clip,keepaspectratio]{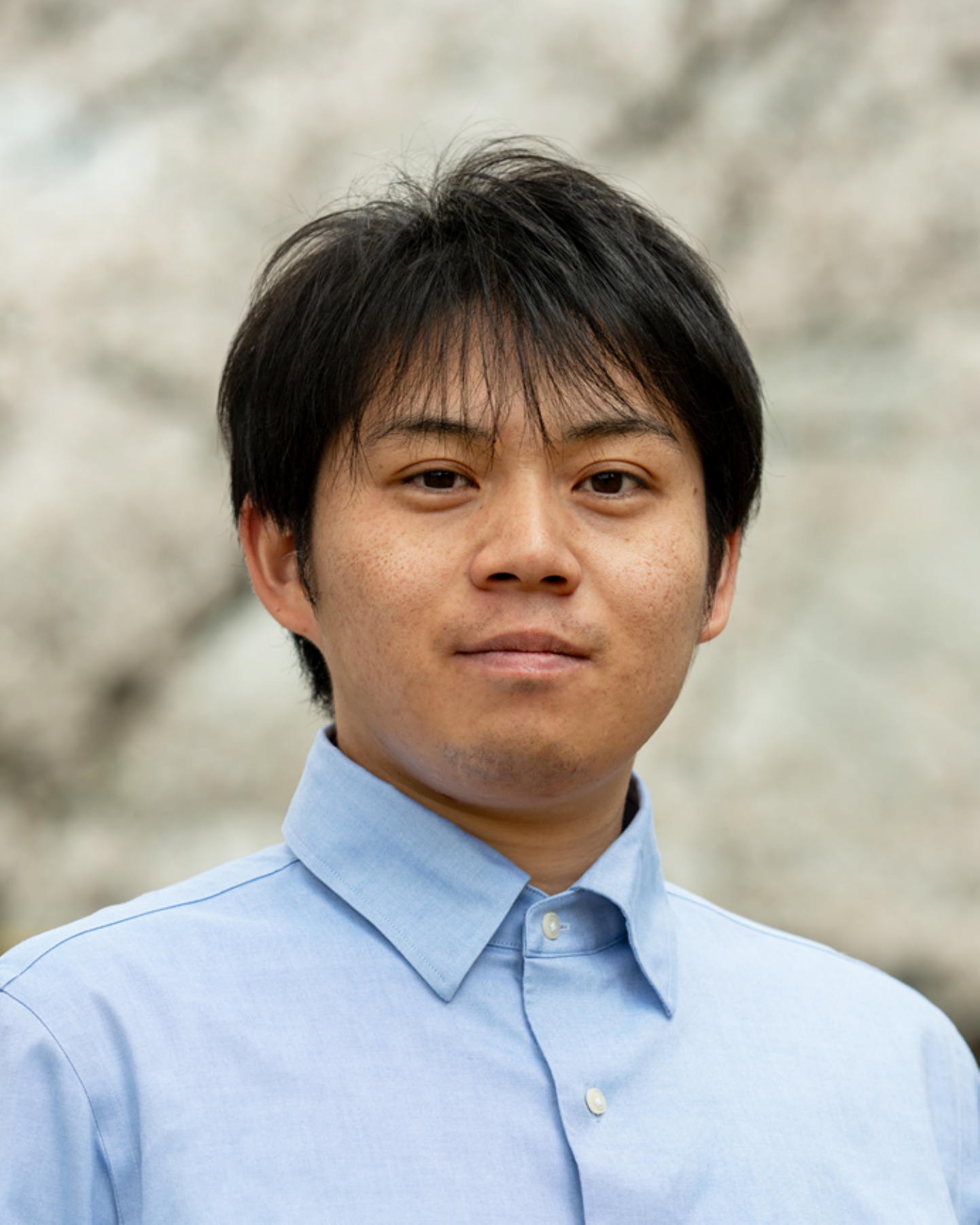}}]{Kazuki Naganuma}
    \spaceskip=.35em plus .15em minus .12em\relax
    (Member, IEEE) received the B.E. degree from Kanagawa Institute of Technology, Atsugi, Japan, in 2020, and the M.E. and Ph.D. degrees from Tokyo Institute of Technology, Meguro, Japan, in 2022 and 2024, respectively, all in information and computer sciences.
    He is an Assistant Professor with the Institute of Engineering, Tokyo University of Agriculture and Technology, Fuchu, Japan. 
    From 2023 to 2025, he was a Research Fellow (DC2) with Japan Society for the Promotion of Science (JSPS), Tokyo, Japan. 
    Since 2025, he has been a Research Fellow (PD) with JSPS, and since 2023, a Researcher with ACT-X, Japan Science and Technology Agency (JST), Tokyo.
    His research interests include signal and image processing and optimization theory.
    He received the Student Conference Paper Award from the IEEE SPS Japan Chapter in 2022, the 38th TELECOM System Technology Student Award from the Telecommunications Advancement Foundation, and the Best Paper Award at APSIPA ASC 2024.
\end{IEEEbiography}
\end{minipage}\par
\vspace{50.027bp}
\noindent\begin{minipage}[t]{\columnwidth}
\begin{IEEEbiography}[{\includegraphics[width=1in,height=1.25in,clip,keepaspectratio]{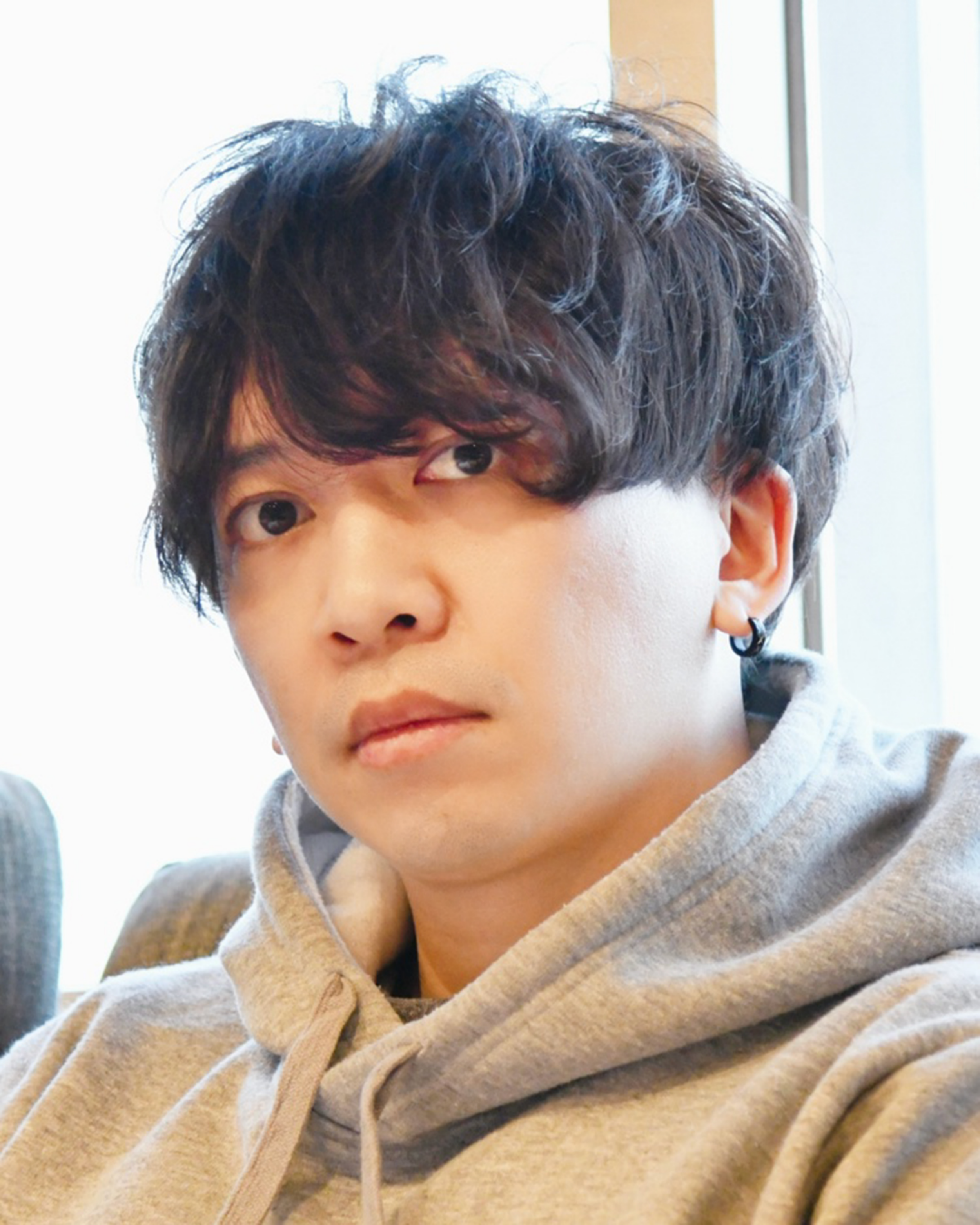}}]{Shunsuke Ono}
    \spaceskip=.35em plus .15em minus .12em\relax
    (Senior Member, IEEE) received the B.E. degree in computer science and the M.E. and Ph.D. degrees in communications and computer engineering from Tokyo Institute of Technology in 2010, 2012 and 2014, respectively. From 2012 to 2014, he was a Research Fellow (DC1) with Japan Society for the Promotion of Science (JSPS). From 2014 to 2024, he was an Assistant Professor and then an Associate Professor with Tokyo Institute of Technology (TokyoTech), Tokyo, Japan. From 2016 to 2020, he was a Researcher with the Precursory Research for Embryonic Science and Technology (PRESTO), Japan Science and Technology Agency (JST), Tokyo. Following the institutional merger in 2024 that established the Institute of Science Tokyo (Science Tokyo), he continued as an Associate Professor until 2026 and has been a Professor since then. His research interests include signal processing, image analysis, optimization, remote sensing, and measurement informatics. He has served as an Associate Editor for IEEE TRANSACTIONS ON SIGNAL AND INFORMATION PROCESSING OVER NETWORKS (2019--2024). He received the Young Researchers' Award and the Excellent Paper Award from the IEICE in 2013 and 2014, respectively, the Outstanding Student Journal Paper Award and the Young Author Best Paper Award from the IEEE SPS Japan Chapter in 2014 and 2020, respectively, and the Best Paper Award at APSIPA ASC 2024. He also received the Funai Research Award in 2017, the Ando Incentive Prize in 2021, the MEXT Young Scientists' Award in 2022, the IEEE SPS Outstanding Editorial Board Member Award in 2023, the KDDI Foundation Award in 2025, and the Special Award for Science Tokyo Advanced Researchers (STAR) in 2026.
\end{IEEEbiography}
\end{minipage}\par

\end{document}